\documentclass[a4paper, 12pt]{article}
\usepackage[T1]{fontenc}
\usepackage{jheppub,graphicx,epsf,amssymb,amsbsy,amsfonts,amssymb,amsmath, empheq,mathrsfs,appendix,soul}
\usepackage{hyperref}

\def\be{\begin{equation}}
\def\ee{\end{equation}}
\def\bea{\begin{eqnarray}}
\def\eea{\end{eqnarray}}

\def\a{\alpha}
\def\b{\beta}
\def\g{\gamma} 
\def\d{\delta}
\def\e{\epsilon}
\def\f{\phi}

\def\l{\lambda}
\def\s{\sigma}\def\l{\lambda}

\def\bg{\bar{g}}
\def\beq{\begin{eqnarray}}\def\eeq{\end{eqnarray}}
\def\ba#1\ea{\begin{align}#1\end{align}}
\def\bg#1\eg{\begin{gather}#1\end{gather}}
\def\bm#1\em{\begin{multline}#1\end{multline}}
\def\bmd#1\emd{\begin{multlined}#1\end{multlined}}

\def\a{\alpha}
\def\b{\beta}

\def\d{\delta}
\def\D{\Delta}
\def\e{\epsilon}

\def\g{\gamma}
\def\G{\Gamma}

\def\l{\lambda}

\def\s{\sigma}

\def\pa{\partial}

\def\({\left(}
\def\){\right)}
\def\[{\left[}
\def\]{\right]}

\def\a{\alpha}

\def\l{\lambda}
\def\b{\beta}
\def\g{\gamma}
\def\d{\delta}
\def\s{\sigma} 
\def\om{\omega}

\def\f{\frac}

\def\D{\Delta}

\title{Singularity Structure of the Four Point Celestial Leaf Amplitudes}

\author{Raju Mandal$^{\,a,c}$, Sagnik Misra$^{\,a,c}$, Partha Paul$^{\,b,c}$ and Baishali Roy$^{\,d} $}
\affiliation[a]{National Institute of Science Education and Research (NISER), Bhubaneswar 752050, Odisha, India}
\affiliation[b]{The Institute of Mathematical Sciences, IV Cross Road, CIT Campus, Taramani, Chennai, India
600113.}
\affiliation[c]{Homi Bhabha National Institute, Training School Complex, Anushakti Nagar, Mumbai 400094, India.}
\affiliation[d]{Ramakrishna Mission Vivekananda Educational and Research Institute, Belur Math, Howrah-711202, West Bengal, India}
\emailAdd{rajuphys002@gmail.com}
\emailAdd{sagnik.misra@niser.ac.in}
\emailAdd{pl.partha13@gmail.com}
\emailAdd{baishali.roy025@gm.rkmvu.ac.in}

\abstract{
In this paper, we study the four-point celestial leaf amplitudes of massless scalar and MHV gluon scattering. These leaf amplitudes are non-distributional decompositions of the celestial amplitudes associated with a hyperbolic foliation of the Klein spacetime. Bulk scale invariance imposes constraints on the total conformal weights of the massless scalars or gluons.
Using this constraint we show that the four-point leaf amplitudes have a \textit {simple pole singularity at $ z = \bar z $}, where, $ z,\bar z $ are two real independent conformal cross ratios. The distributional nature of the four-point celestial amplitudes is recovered by adding the leaf amplitudes in the timelike and spacelike wedges of the spacetime.
We also verify that the MHV gluon leaf amplitudes satisfy a set of differential equations previously obtained for celestial MHV gluon amplitudes by considering the soft gluon theorems and the subleading terms in the OPE expansion between two positive helicity gluons.}

\begin{document}
\maketitle
\flushbottom

\section{Introduction}

Celestial holography is a proposed duality between quantum gravity in four-dimensional (4D) asymptotically flat spacetime and a two-dimensional (2D) conformal field theory (CFT) living on the celestial sphere \cite{Strominger:2017zoo,Pasterski:2021rjz,Donnay:2023mrd,Pasterski:2016qvg}. The central objects of study in celestial holography are the celestial amplitudes. They are $ S $-matrix elements written in boost eigenbasis rather than usual momentum eigenbasis. For massless scattering, they are obtained from the momentum space scattering amplitudes by Mellin integrating the energies of the external massless particles \cite{Pasterski:2017kqt,Banerjee:2018gce}. Celestial amplitudes transform as  2D conformal correlators under global (Lorentz) conformal transformations. Thus we can apply the powerful 2D CFT techniques to constrain the quantum gravity scattering amplitudes. However, one of the caveats about this approach is that, lower point celestial amplitudes are heavily constrained due to the spacetime translation symmetries. They take distributional forms which are unfamiliar from the usual 2D CFT perspective. Several efforts have been made to construct smooth conformally invariant celestial amplitudes by breaking translational symmetries \cite{Costello:2022wso,Fan:2022vbz,Casali:2022fro,Gonzo:2022tjm,Stieberger:2023fju,Costello:2022jpg,Stieberger:2022zyk,Melton:2024akx,Bittleston:2023bzp,Costello:2023hmi,Adamo:2023zeh,Ball:2023ukj,Banerjee:2023rni}.

Recently, authors in \cite{Melton:2023bjw}, have considered three-point MHV gluon amplitudes in $ (2,2) $ signature Klein spacetime and showed that the translationally invariant celestial amplitudes can be written as sums of generically smooth amplitudes given by $ \text{AdS}_3/\mathbb{Z} $-Witten diagrams. These amplitudes are called  \textit{leaf amplitudes}. Klein space can be divided into  timelike ($ X^2<0 $) and spacelike ($ X^2>0 $) wedges, each of which is foliated by hyperbolic slices. These slices are geometrically $ \text{AdS}_3/\mathbb{Z} $ whose boundaries are Lorentzian torus.  Using the Fourier representation of the momentum-conserving delta function, celestial amplitudes can be written directly in position space as the weighted integrals of Witten diagrams on these slices/leaves. These leaf amplitudes enjoy Lorentz/conformal symmetry but they are not translationally invariant. Thus they take the familiar 2D CFT form on the Lorentzian torus. For a discussion on 2D CFT correlators on Lorentzian torus see \cite{Melton:2023hiq}. 

In this paper, we study the four-point leaf amplitudes of massless scalar and MHV gluon scattering \footnote{Since two- and three-point functions are completely fixed by symmetries, they don't provide any information about the bulk dynamics.}. In general these four-point leaf amplitudes have branch point singularities as a function of two real independent cross ratios $ z $ and $ \bar z $. The total conformal weights of the massless scalars or gluons are constrained since the bulk theories under consideration are scale invariant. On the support of these constraints, we compute the discontinuities around the branch point singularities and show that the four-point leaf amplitudes are non vanishing everywhere on the cross-ratio space defined by $ z, \bar z $ and develop a \textit {simple pole singularity at $ z = \bar z $}. The distributional nature of the four point celestial amplitudes will be recovered by adding the leaf amplitudes in both timelike and spacelike regions.



Interestingly, similar singularity structure of the four-point boundary correlators have appeared in other studies also \cite{Jain:2023fxc,Alday:2024yyj,Gary:2009ae,Maldacena:2015iua}. Recently, in \cite{Jain:2023fxc}, an interesting approach  has been discussed to study the flat space holography \footnote{See \cite{Banerjee:2024yir} for a connection between their approach and celestial holography.}. More specifically, by considering the Euclidean path integral of a quantum field theory (QFT) the authors of \cite{Jain:2023fxc} have defined  boundary correlation functions with Dirichlet boundary conditions. The $ S $-matrix of the QFT can be obtained by smearing these boundary correlators with the free particle wave functions. In the case of massless scalar scattering, after analytically continuing to Minkowski spacetime they have found that their four-point boundary correlators develop a simple pole singularity at $ z = \bar z $, where $ z, \bar z $ are the conformal cross ratios on the celestial sphere. In the AdS/CFT context, on the other hand,  it was shown that there exists certain singularity structure for the four-point Lorentzian boundary correlators, known as the bulk point singularity and this is directly related to the bulk locality in \cite{Gary:2009ae,Maldacena:2015iua}. The residues of these singularities give the flat space $ S $-matrix elements. However, one needs to consider certain limits to get the $ S $-matrix elements from the boundary correlators.


%

In our analysis, we have considered two examples: massless scalar and MHV gluon scattering and, have shown that the four point leaf amplitudes correspond to simple pole singularity on the support of the delta function constraint involving total conformal weights. It will be interesting to see if our conclusion holds without this constraint as well. Another interesting question is to check the singularity structure of the four point leaf amplitudes corresponding other bulk scattering. We hope to answer some these questions in future.


In this paper, we also show that the leaf amplitudes satisfy the Banerjee-Ghosh (BG) equations first derived in \cite{Banerjee:2020vnt} for MHV gluon celestial amplitudes. By considering the ward identities of leading \cite{Strominger:2014ass,Fan:2019emx,Donnay:2018neh,Pate:2019mfs,Nandan:2019jas,Adamo:2019ipt,He:2014cra,Kapec:2015ena,Kapec:2014zla,Campiglia:2015qka,Nande:2017dba,He:2020ifr} and subleading \cite{Banerjee:2020vnt,Guevara:2021abz,Strominger:2021lvk,Strominger:2021mtt} positive helicity soft gluon theorems and the OPE between two outgoing positive helicity gluons, the authors in \cite{Banerjee:2020vnt}, found some null state relation under the leading and subleading soft gluon algebra. Decoupling of these null states from the MHV gluon celestial amplitudes leads to differential equations for the latter. Later, this was generalized to the whole $ S $-algebra and it was found that there exists an (discrete) infinite number of theories invariant under $ S $-algebra, each having a null state relation \cite{Banerjee:2023bni}. Like celestial MHV gluon amplitudes, leaf amplitudes are also governed by the same infinite dimensional soft gluon algebra \cite{Melton:2024jyq}, known as $ S $-algebra. Thus, it is expected that they will satisfy the BG equations. By computing the subleading terms in the OPE between two outgoing positive helicity hard gluon primary operators and comparing it to the OPE between a soft and a hard gluon primary operator, we derive the BG equations for leaf amplitudes.

The rest of the paper is organised as follows. We summarize our main results in the next section. To make the paper self consistent, in section \ref{revw}, we give a brief review of the geometry of Klein space and the construction of the leaf amplitudes. Section \ref{sing_sec} discusses the singular behavior of the four-point leaf amplitudes in the cross ratio space. We consider two examples: scalar contact diagram and MHV gluon amplitudes. In section 
\ref{BG_eq}, by extracting the subleading order $ (\mathcal{O}(1)) $ OPE from the four-point MHV gluon leaf amplitudes we show that  they satisfy the BG equations. Appendix \ref{appndxlc} computes the integral appeared in the four-point scalar leaf amplitude in detail. In Appendix \ref{H_func}, we discuss some of the properties of $ H $-function used in the main section. Finally in Appendix \ref{scal_cel_amp} we have computed the tree level celestial four-point scalar amplitude to match with the results obtained in subsection \ref{recov}. Let us now start with the summary of the main results.

\section{Summary of the main results}

We denote the conformally invariant part of the $ 4 $-point celestial leaf amplitudes for tree level massless scalar scattering by  $ \mathcal{S}_4(z, \bar z) $, where $ z $ and $ \bar z $ are two real conformally invariant cross ratios. This is the leaf amplitude in the timelike region of the Klein space. Similarly, for spacelike region we have $ \overline{\mathcal{S}_4}(z, \bar z) $. The full leaf amplitudes will be obtained by multiplying the appropriate conformally covariant (non-unique) pre-factor. On the support of the constraints on the imaginary part $ \b $, of the total conformal weights  coming from the bulk scale invariance, these leaf amplitudes develop a simple pole singularity at $ z = \bar z $. We know that the conformal dimension $ \D_i $ of the $ i $-th primary operator on the celestial torus lies on the principle continuous series, i.e., $ \D_i = 1+i\l_i $ \cite{Pasterski:2017kqt}. For simplicity we take, $ \l_i = \l, \forall i $. Then $ \b $ becomes $ \b=4\l $. Then our main results for $ 4 $-point scalar leaf amplitudes are given by,
\be 
\begin{split}
 \d(\l)\mathcal{S}_4(z, \bar z)\big|_{z \to \bar z} &= \d(\l)\f{i\pi}{2} \f{4\pi^2}{\bar z - z + i\e}, \ z>1 \\
 \d(\l)\overline{\mathcal{S}_4}(z, \bar z)\big|_{z \to \bar z} &= -\d(\l)\f{i\pi}{2} \f{4\pi^2}{\bar z - z - i\e}, \ z>1.
\end{split}
\ee
The celestial amplitudes will be obtained by adding the two leaf amplitudes on the support of $ \d(\l) $ and we will recover the distributional nature of the celestial amplitudes as explained in detail in section \ref{sc_cn_diag}.

Similarly, for MHV gluon scattering we define $ \mathcal{G}_4(z,\bar z) $ and $ \overline{\mathcal{G}_4}(z,\bar z) $ the two conformally invariant part of the $ 4 $-point leaf amplitudes in timelike and spacelike regions respectively. We then have,
\be 
\begin{split}
 \d(\l)\mathcal{G}_4(z, \bar z)\big|_{z \to \bar z} &= \d(\l)\f{i\pi}{2} \f{z}{z-1} \f{4\pi^2}{\bar z - z + i\e}, \ z>1 \\
 \d(\l)\overline{\mathcal{G}_4}(z, \bar z)\big|_{z \to \bar z} &= -\d(\l)\f{i\pi}{2} \f{z}{z-1} \f{4\pi^2}{\bar z - z - i\e}, \ z>1.
\end{split}
\ee

Once again the addition of the above two leaf amplitudes will give the distributional nature of the conformally invariant part of the $ 4 $-point celestial MHV gluon amplitude. For details see \ref{mhv_gluon_scatt}.

\section{Celestial leaf amplitudes in Klein space}
\label{revw}

In this section, we first briefly review the geometry of Klein space. For details one can see \cite{Atanasov:2021oyu,Mason:2005lj}. Then, following \cite{Melton:2023bjw}, we describe the construction of celestial leaf amplitudes in Klein space.

\subsection{The geometry of Klein space}

Klein space, denoted as $ \mathbb{K}^{2,2} $, is a (2,2)-signature flat space with the metric in Cartesian coordinates given by,
\be \label{metric_cart}
ds^2 = -(dX^0)^2 - (dX^1)^2 + (dX^2)^2 + (dX^3)^2
\ee 
We want to conformally compactify $ \mathbb{K}^{2,2} $ and study the conformal geometry of null infinity, denoted as $ \mathcal{I} $.
In polar coordinates,
\be
X^0+iX^1 = q e^{i\psi}, \ X^2 + iX^3 = re^{i\phi} 
\ee
the metric becomes
\be \label{polar_metric}
ds^2 = - dq^2 - q^2 d\psi^2  + dr^2 + r^2 d\phi^2  
\ee

To study the null infinity $ \mathcal{I} $, we define the following coordinates,
\be
q-r = \tan U, \ q+r = \tan V 
\ee

Since $ 0< q,r < \infty $, the coordinate ranges for $ (U, V) $ are $ -\f{\pi}{2} < U < \f{\pi}{2}, \ |U| < V < \f{\pi}{2} $. In these $ (U,V) $ coordinates, the metric \eqref{polar_metric} becomes,
\be\label{metric_UV}
ds^2 = \f{1}{\cos^2 U \cos^2 V}\( -dU dV - \f{1}{4}\sin^2(V+U)d\psi^2 + \f{1}{4}\sin^2(V-U)d\phi^2 \) 
\ee

The timelike infinity $ (i') $ is obtained by sending $ q \to \infty $ where $ \(U=\f{\pi}{2}, V = \f{\pi}{2}\) $ and spacelike infinity is given by $ r \to \infty $, where $ \(U=-\f{\pi}{2}, V = \f{\pi}{2}\)  $. Null infinity is at $ V = \f{\pi}{2}. $ It is parametrized by the null coordinate $ \ -\f{\pi}{2} < U < \f{\pi}{2} $ and the periodic coordinates $ (\psi, \phi) $. Taking the limit $ V \to \f{\pi}{2} $ and rescaling the metric \eqref{metric_UV} by $ \cos^2 V $, we get the conformal metric on $ \mathcal{I} $ as given by,
\be \label{metric_I}
ds^2 = -d\psi^2 + d\phi^2, \qquad \psi \sim \psi + 2\pi, \ \phi \sim \phi + 2\pi. 
\ee


 Thus we see that, in this signature the null infinity is given by Lorentzian torus ($\mathcal{LT}^2 = S^1 \times S^1 $) times a null interval. It has a single connected component and we can not define incoming and outgoing particles both. Hence, unlike Minkowski space $ (\mathbb{M}^{1,3}) $, we can't define an $ S $-matrix, rather we have an $ S $-vector. However, a suitable analytic continuation can be used to define an $ S $-matrix in $ \mathbb{M}^{1,3} $ from $ S $-vector in $ \mathbb{K}^{2,2} $ \footnote{See \cite{Melton:2024pre} for more recent discussion on this.}.

 
 The lightcone of the origin in Cartesian coordinates is given by,
\be
X^2 = 0 
\ee

\begin{figure}
\begin{center}
\includegraphics[height=8cm,width=9cm]{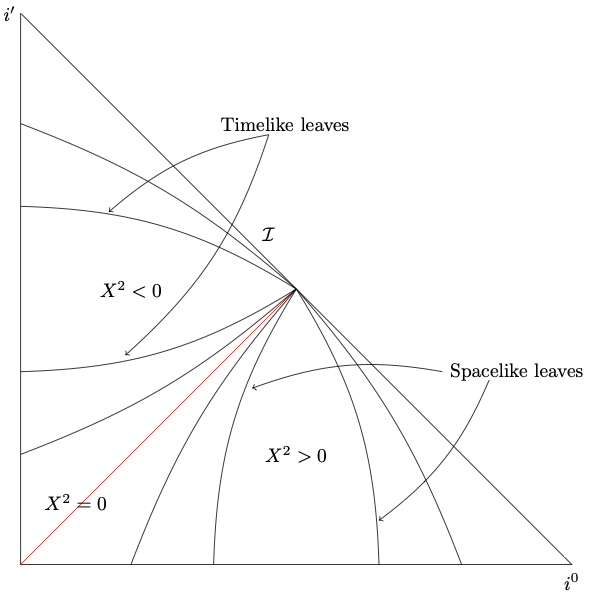}
\caption{Penrose diagram of Klein space $ \mathbb{K}^{2,2} $ where the hyperbolic foliation in timelike and spacelike region is shown. Each point on this diagram denotes a Lorentzian torus.}
\label{KS:fig1}
\end{center}
\end{figure}

It divides the Klein space into two regions: timelike $ X^2 < 0 $ and spacelike $ X^2 > 0 $ (See figure \ref{KS:fig1}). In both the regions we can parametrize the coordinates as follows:
\be \label{tdst}
\begin{split}
\textnormal{Timelike}&: X^\mu = \tau \hat{x}^\mu_+, \  \hat{x}^2_+ = -1 \\
\textnormal{Spacelike}&: X^\mu = \tau \hat{x}^\mu_-, \  \hat{x}^2_- = + 1.
\end{split} 
\ee
where $ \tau \in (0,\infty) $ is the magnitude of $ X^\mu $. Constant $ \tau $ surfaces on both the regions constitute the hyperbolic foliation of $ \mathbb{K}^{2,2} $, similar to the hyperoblic foliation of Minkowski space \cite{deBoer:2003vf}. We now study the geometry of constant $ \tau $ slices or leaves. The timelike leaves $ \hat x^2_{+} = -1 $ can be parametrized using the global coordinates,
\be \label{smallx}
\hat{x}^0_+ + i \hat{x}^1_+ = e^{i\psi} \cosh\rho, \ \hat{x}^2_+ + i \hat{x}^3_+ = e^{i\phi} \sinh\rho 
\ee 

where $ 0< \rho < \infty $ is radial coordinate along each of the constant $ \tau $-slices. In these coordinates metric \eqref{metric_cart} becomes,
\be\label{time-like-metric}
ds^2 =  - d\tau^2 + \tau^2 \(-\cosh^2 \rho \, d\psi^2 + \sinh^2 \rho \, d\phi^2  + d\rho^2 \)
\ee

Thus each constant $ \tau $ leaf gives a metric on $ \textnormal{AdS}_3$ with a periodically identified time coordinate $ \psi $, hence will be denoted as $ \textnormal{AdS}_3/\mathbb{Z} $. Similar description holds for spacelike constant $ \tau $-slices also, except that the role of timelike and spacelike cycles now exchanged:
\be
 \hat{x}^0_- + i \hat{x}^1_- = e^{i\psi} \sinh\rho, \ \hat{x}^2_- + i \hat{x}^3_- = e^{i\phi} \cosh\rho 
\ee
and the metric in the spacelike region is given by,
\be\label{space-like-metric}
ds^2 =  d\tau^2 - \tau^2 \(-\cosh^2 \rho \, d\phi^2 + \sinh^2 \rho \, d\psi^2  + d\rho^2 \)
\ee

The conformal boundary of each $ \textnormal{AdS}_3/\mathbb{Z} $ leaf \eqref{time-like-metric} or \eqref{space-like-metric} can be obtained by sending $ \rho \to \infty $ and is given by the Lorentzian torus \eqref{metric_I}. We call the Lorentzian torus on $ \mathcal{I} $ as celestial torus $ \mathcal{C T}^2 $.

We can introduce the null coordinates (also called the global coordinates) 
\be \label{null_coord}
\s = \f{\psi+\phi}{2}, \ \bar \s = \f{\psi-\phi}{2} 
\ee
on $ \mathcal{C T}^2 $. It is convenient to compute the leaf amplitudes using these global coordinates and then for our purpose we will convert the results to the planar coordinates, defined as,
\be \label{planar_coord}
z = \tan\s, \ \bar z = \tan\bar\s, \qquad z,\bar z \in \mathbb{R}
\ee 

The Lorentz transformations $ (\textnormal{SL}(2,\mathbb{R}) \times \overline{\textnormal{SL}}(2,\mathbb{R})) $ act on these planar coordinates as real independent Mobius transformations
\be
\begin{split}
z &\to \f{a z + b}{c z + d}, \qquad (a,b,c,d) \in \mathbb{R}, \qquad ad-bc=1. \\
\bar z &\to \f{\bar a \bar z + \bar b}{\bar c \bar z + \bar d}, \qquad (\bar a, \bar b, \bar c, \bar d) \in \mathbb{R}, \qquad \bar a \bar d- \bar b\bar c=1.
\end{split} 
\ee

From the definition \eqref{planar_coord} it is clear that the planar coordinates $ (z, \bar z) $ can not differentiate between the points $ (\psi, \phi) $ and $ (\psi+\pi, \phi+\pi) $ and, hence they cover only half of the celestial torus. Rather, the coordinates $ (z, \bar z) $ can be understood as the local coordinates of a two dimensional diamond in $ (1,1) $ signature and the whole celestial torus except the point $ z \bar z = 0 $ can be covered by two such diamonds. However, it should be noted that the distinction between two diamonds is just a choice of the coordinate patches, and in the global coordinates $ (\s,\bar \s) $ we don't have any such distinction.

\subsection{Leaf amplitudes}

Let us first introduce our notations. For any null momentum $ p_k^\mu $ of $ k $-th massless particle, satisfying $ p_k^2=0 $, we use the following parametrization in planar coordinates in $ (-,-,+,+) $ signature:
\be \label{pmu}
\begin{gathered}
p_k^\mu = \e_k \om_k q_k^\mu, \ q_k^\mu = \(1-z_k \bar z_k, z_k +\bar z_k, 1+z_k \bar z_k, z_k -\bar z_k\)
\end{gathered} 
\ee

where $ \om_k \in [0,\infty) $ represents the magnitude of the frequency of the $ k $-th massless particle and $ \e_k = \pm $ denotes the sign of the frequency. Momentum \eqref{pmu} can also be parametrized in terms of global coordinates $ (\s, \bar \s) $ through the relation \eqref{planar_coord} and \eqref{null_coord},
\be\label{mom_gc}
p_k^\mu = \f{\om_k \hat{p}_k^\mu }{|\cos\s_k \cos\s_k|} 
\ee
where $ \hat{p}_k^\mu $ is a null vector parametrized by the points $ (\psi_k, \phi_k) $ on $ \mathcal{C T}^2 $ as,
\be \label{phat}
\hat{p}_k^0 + i\hat{p}_k^1=e^{i\psi_k}, \ \hat{p}_k^2 +i\hat{p}_k^3=e^{i\phi_k}
\ee

and we have identified $ \e_k = \f{p_k^0 + p_k^2}{2\om_k} = \textnormal{sgn}(\cos\s_k \cos\bar\s_k) $.

We start with an $ n $-point tree level momentum space amplitude with all the external state particles as massless,
\be \label{msamp}
\begin{gathered}
A_n(1^{\vartheta_1},2^{\vartheta_2},\ldots, n^{\vartheta_n}) = A_n(\{\e_i\om_i q_i,\vartheta_i\}) \, \d\(\sum_{k=1}^n \e_k \om_k q_k^\mu\) 
\end{gathered}
\ee
where $ (\vartheta_1,\vartheta_2, \cdots, \vartheta_n) $ are the helicities of the massless particles and we have written the momentum-conserving delta function explicitly. The celestial amplitude is obtained by performing Mellin transformation with respect to each external massless leg,
\be\label{melamp0}
\begin{gathered}
\mathcal{M}_n\(1^{(h_1,\bar h_1)},2^{(h_2,\bar h_2)}, \ldots, n^{(h_n,\bar h_n)}\) = \prod_{j=1}^n \int_0^\infty d\om \, \om^{\D_j-1} e^{-\e \om_j}A_n(1^{\vartheta_1},2^{\vartheta_2},\ldots, n^{\vartheta_n})
\end{gathered} 
\ee
where $ \e $ is a UV regulator and $ h_j, \bar h_j $ are conformal weights given by,
\be
h_j=\f{\D_j+\vartheta_j}{2}, \ \bar h_j=\f{\D_j-\vartheta_j}{2}.
\ee
Then we use the Fourier representation of the delta function
\be
\d\(\sum_{k=1}^n \e_k \om_k q_k^\mu\) = \int \f{d^4 X}{(2\pi)^4}  e^{i \sum_{k=1}^n \e_k \om_k q_k \cdot X }
\ee
to write the amplitude \eqref{melamp0} as,
\be\label{melamp}
\begin{gathered}
\mathcal{M}_n\(1^{(h_1,\bar h_1)},2^{(h_2,\bar h_2)}, \ldots, n^{(h_n,\bar h_n)}\)  = \prod_{j=1}^n \int_0^\infty d\om_j \, \om_j^{\D_j-1} e^{-\e \om_j} \\
\times A_n(\{\e_i\om_i q_i, \vartheta_i \})  \int \f{d^4 X}{(2\pi)^4}  e^{i \sum_{k=1}^n \e_k \om_k q_k \cdot X }
\end{gathered} 
\ee

Mellin integral of the plane wave $ e^{i  \e_k \om_k q_k \cdot X } $ gives the scalar conformal primary wavefunction $ \Phi_{\D_k,\e_k}(X, q_k) $ given by \eqref{scal_conf_prim}. For the kind of amplitudes that we are interested in, the Mellin integral with respect to the energies $ \om_j $ in \eqref{melamp} can be performed and the resultant expression can be written as the spacetime integral of $ \Phi_{\D_k,\e_k}(X, q_k) $'s  \cite{Melton:2023bjw}. Hence, the amplitudes on the LHS of \eqref{melamp} can be thought of as evaluated directly in the position space without referring to momentum space amplitudes. However, we will keep the momentum space form of the amplitudes in \eqref{melamp} for our convenience.

Now, dividing the $ X^\mu $ integral in \eqref{melamp}  into two parts according to \eqref{tdst} and replacing $ \om_i \to \f{\om_i}{\tau} $ we get,
\be\label{main_int}
\begin{gathered}
\mathcal{M}_n\(1^{(h_1,\bar h_1)},2^{(h_2,\bar h_2)}, \ldots, n^{(h_n,\bar h_n)}\)  = \f{1}{(2\pi)^4}\int_0^\infty \tau^{3-n-\b} d\tau \[ \int_{\hat{x}_+^2=-1}  d^3 \hat{x}_+ \prod_{j=1}^n \int_0^\infty d\om_j \, \om_j^{\D_j-1} \right. \\
\left. \times A_n\(\left\{\f{\e_i\om_i q_i}{\tau},\vartheta_i \right\}\)  e^{i \sum_{k=1}^n \e_k \om_k q_k \cdot \hat{x}_+ -\e \om_k} + \int_{\hat{x}_+^2=1}  d^3 \hat{x}_- \prod_{j=1}^n \int_0^\infty d\om_j \, \om_j^{\D_j-1} A_n\(\left\{\f{\e_i\om_i q_i}{\tau},\vartheta_i \right\}\) \right.\\
\left. \times e^{i \sum_{k=1}^n \e_k \om_k q_k \cdot \hat{x}_- -\e \om_k} \]
\end{gathered} 
\ee

where $ \b = \sum_{k=1}^n (\D_k-1) = \sum_{k=1}^n (2\bar h_k+\vartheta_k-1) $. The scaling behavior of an $ n $-point scattering amplitude without the momentum conserving delta function in any scale invariant theory is given by
\be \label{scaling_behavior}
\begin{gathered}
A_n\(\left\{\f{\e_i\om_i q_i}{\tau},\vartheta_i \right\}\) 
\to \tau^{n-4} A_n\(\left\{\e_i\om_i q_i,\vartheta_i \right\}\)
\end{gathered}
\ee
This can be seen from the mass dimension of the $ n $-point amplitude. For an $ n $-point scalar or gluon amplitudes the mass dimension is $ [A_n] \sim (\text{mass})^{4-n} $ and hence the scaling behavior is given by \eqref{scaling_behavior} \cite{Gross:1970tb}.  With this scaling behavior, the $ \tau $ integral in \eqref{main_int} can be performed producing $ \d(\b) $ and the resulting expression can be written as
\be\label{cel_amp}
\mathcal{M}_n\(1^{(h_1,\bar h_1)},2^{(h_2,\bar h_2)}, \ldots, n^{(h_n,\bar h_n)}\)  = \f{\d(\b)}{(2\pi)^3} \[\mathcal{L}_n(\s_i,\bar \s_i) + \overline{\mathcal{L}}_n(\s_i,\bar \s_i) \]
\ee
where
\be \label{la1}
\begin{split}
& \hspace{2cm} \mathcal{L}_n(\s_i,\bar \s_i) = \int_{\hat{x}_+^2=-1}  d^3 \hat{x}_+ \prod_{j=1}^n \int_0^\infty d\om_j \, \om_j^{\D_j-1} A_n\(\left\{\e_i \om_i q_i,\vartheta_i \right\}\) e^{i \sum_{k=1}^n \e_k \om_k q_k \cdot \hat{x}_+ -\e \om_k} \\
& \hspace{2cm} \overline{\mathcal{L}}_n(\s_i,\bar \s_i) = \int_{\hat{x}_+^2=1}  d^3 \hat{x}_- \prod_{j=1}^n \int_0^\infty d\om_j \, \om_j^{\D_j-1} A_n\(\left\{\e_i\om_i q_i,\vartheta_i\right\} \) e^{i \sum_{k=1}^n \e_k \om_k q_k \cdot \hat{x}_- -\e \om_k}
\end{split} 
\ee

The above amplitudes $ \mathcal{L}_n $ and $ \overline{\mathcal{L}}_n $ are called $ n $-point celestial leaf amplitudes. We have shown their dependence on the global coordinates explicitly, however it should be implicitly understood that these amplitudes also depend on the conformal weights which have been suppressed for notational convenience. Additionally, a change of variables shows that the second integral in \eqref{la1} can be obtained from the first one by sending $ \bar \s_i \to -\bar \s_i $ i.e., 
\be
\overline{\mathcal{L}}_n(\s_i,\bar \s_i) = \mathcal{L}_n(\s_i, - \bar \s_i)
\ee 
We will now focus on $ 4 $-point celestial leaf amplitudes for scalars and gluons in the next section and analyze their singularity behavior in cross ratio space on the support of $ \d(\b) $.

\section{Singularity structure of the four-point leaf amplitudes on the support of $ \d(\b) $}
\label{sing_sec}

In the previous section we have discussed how, given a momentum space amplitude, one can construct its corresponding leaf amplitudes by performing the integrals \eqref{la1}. In this section we will first construct the four-point leaf amplitudes in planar coordinates for tree level scalar and MHV gluon scattering and then show that each leaf amplitude contains a simple pole at $ z = \bar z $ on the support of $ \d(\b) $.

\subsection{Scalar contact diagram}
\label{sc_cn_diag}

Let us consider the massless $ \phi^4 $ theory. At tree level we only have the contact diagram and the amplitude is given by
\be  \label{s4pt}
A_4(p_1,p_2,p_3,p_4) = - i (2\pi)^4 \tilde{\l} \, \d^{(4)}(p_1+p_2+p_3+p_4)
\ee 
where $ \tilde{\l} $ is the coupling constant. From \eqref{msamp}, we then have
\be \label{tlsa}
A_4\(\left\{\e_i \om_i q_i , \vartheta_i \right\}\)  = - i (2\pi)^4 \tilde{\l}
\ee

Hence, in this case from \eqref{la1} the 4-point leaf amplitudes are given by,
\be \label{lasc}
\begin{split}
\mathcal{L}^s_4(\s_i, \bar \s_i) &= - i (2\pi)^4 \tilde{\l} \ \mathcal{C}_4 (\s_i, \bar \s_i), \\  \overline{\mathcal{L}^s}_4(\s_i, \bar \s_i) &= \mathcal{L}^s_4(\s_i, - \bar \s_i)  = - i (2\pi)^4 \tilde{\l} \  \mathcal{C}_4 (\s_i, -\bar \s_i) 
\end{split} 
\ee

where 
\be \label{gen_c4}
\begin{split}
\mathcal{C}_4 (\s_i, \bar \s_i) &= \int_{\hat{x}_+^2=-1}  d^3 \hat{x}_+ \prod_{j=1}^4 \int_0^\infty d\om_j \, \om_j^{2 \bar h_j-1} e^{i \sum_{k=1}^4 \e_k \om_k q_k \cdot \hat{x}_+ -\e \om_k} \\
\end{split} 
\ee
and we have introduced a superscript $ s $ on $ \mathcal{L}, \overline{\mathcal{L}} $ to denote scalars. Here we want to emphasize that for scalars we have $ \vartheta_j = 0,  \ \forall j $, which says that the conformal weights $ h_j,\bar  h_j $ are given by $ h_j = \f{\D_j}{2} = \f{1+i\l_j}{2} =\bar  h_j $. However, we will perform the integrals without using any specific values for $ \bar h_j $'s as the integral \eqref{gen_c4} will appear for MHV gluon scattering also. At this stage it is useful to introduce the abbreviation for the torus separation
\be\label{tor_sep}
s_{ij} := \sin(\s_{ij}), \ \bar s_{ij} := \sin (\bar \s_{ij}).
\ee  
where $ \s_{ij} = \s_i - \s_j, \ \bar \s_{ij} = \bar \s_i - \bar \s_j $.

The integral \eqref{gen_c4} has been computed in detail in appendix \ref{appndxlc}. Using the abbreviation $S_{ij}= s_{ij}\bar s_{ij} $, here we write the final result:
\be \label{main_res}
\begin{gathered}
\mathcal{C}_4(\s_i,\bar \s_i) = \f{i\pi}{2} \G\(\bar h - 1\) ( S_{13}+i\e)^{d_1}( S_{34}  + i\e)^{d_2} ( S_{24}+i\e)^{d_3} ( S_{23} + i\e)^{d_4} H\( u_+, v_+\) \\
-\f{i\pi}{2} \G\(\bar h - 1\) ( S_{13} -i \e)^{d_1}( S_{34} -i \e)^{d_2} ( S_{24} -i\e)^{d_3} ( S_{23} -i \e)^{d_4} H\(u_-, v_-\)
\end{gathered} 
\ee

where 
\be 
 d_1=-2\bar h_1, \ d_2 = \bar h_1 +\bar h_2 - \bar h_3 - \bar h_4, \ d_3 = -\bar h_1 - \bar h_2 + \bar h_3 - \bar h_4, \ d_4 = \bar h_1 - \bar h_2 - \bar h_3 + \bar h_4.
\ee
and $ u_{\pm}, v_{\pm} $ are given by
\be 
u_{\pm} = \f{( s_{12} \bar s_{12})( s_{34} \bar s_{34} \pm i\e)}{( s_{13} \bar s_{13} \pm i\e)( s_{24} \bar s_{24} \pm i\e)}, \ v_{\pm} = \f{\( s_{14} \bar s_{14}\)( s_{23} \bar s_{23} \pm i\e)}{( s_{13} \bar s_{13} \pm i\e)( s_{24} \bar s_{24} \pm i\e)}
\ee
 $ H $-functions in \eqref{main_res} are discussed in \cite{Dolan:2000uw} and reviewed in appendix \ref{H_func}. For notational convenience we have suppressed their first four arguments. However, we will show them explicitly whenever it will be required. As we mentioned before, for our purpose it will be useful to write the leaf amplitudes in planar coordinates. The coordinate transformation is,
\be \label{plan_coord_tranf}
s_{ij} = z_{ij} \cos\s_i \cos\s_j, \ \bar s_{ij} = \bar z_{ij} \cos\bar \s_i \cos\bar \s_j, \e_i = \text{sgn}(\cos\s_i \cos\bar \s_i)
\ee

After the coordinate transformations and multiplying by the Jacobian factor $ \prod_{i=1}^4 |\cos\s_i|^{2h_i} |\cos\bar \s_i|^{2\bar h_i} $ we will get the leaf amplitudes in the planar coordinates $ (z_i,\bar z_i) $. For $ \mathcal{C}_4(z_i,\bar z_i) $ these steps give,
\be
\begin{gathered}\label{sa_pc1}
\mathcal{C}_4(z_i,\bar z_i) = \f{i\pi}{2} \G\(\bar h - 1\) ( \e_1 \e_3 z_{13} \bar z_{13} + i \e)^{d_1}(\e_3 \e_4 z_{34} \bar z_{34} + i\e)^{d_2} (\e_2 \e_4 z_{24} \bar z_{24}+i\e)^{d_3} \\
\times (\e_2 \e_3 z_{23} \bar z_{23} + i\e)^{d_4} H\(u_+, v_+\) - \f{i\pi}{2}  \G\(\bar h - 1\) (\e_1 \e_3 z_{13} \bar z_{13} -i \e)^{d_1} \\
\times (\e_3 \e_4 z_{34} \bar z_{34} -i \e)^{d_2} (\e_2 \e_4 z_{24} \bar z_{24}-i\e)^{d_3}(\e_2 \e_3 z_{23} \bar z_{23} -i \e)^{d_4} H\(u_-,v_-\)\\
\end{gathered} 
\ee
 where in planar coordinates $ u_{\pm}, v_{\pm} $ are now given by,
 \be \label{gen_uv}
 u_{\pm} = \f{(\e_1 \e_2 z_{12} \bar z_{12})(\e_3 \e_4 z_{34} \bar z_{34} \pm i \e)}{(\e_1 \e_3 z_{13} \bar z_{13} \pm i \e)(\e_2 \e_4 z_{24} \bar z_{24} \pm i \e)}, v_{\pm} = \f{\( \e_1 \e_4 z_{14} \bar z_{14}\)(\e_2 \e_3 z_{23} \bar z_{23} \pm i\e)}{( \e_1 \e_3 z_{13} \bar z_{13} \pm i\e)(\e_2 \e_4 z_{24} \bar z_{24} \pm i\e)}
 \ee

Here, we want to emphasize one important point that in planar coordinates the signs of the frequencies $ \e_i $ reappear. Because in planar coordinates $ \mathcal{CT}^2 $ gets divided into two diamonds and the signs $ \e_i $ determine the choice of the diamonds. This is equivalent to the choice of the $ \text{in} $ vs. out in Minkowski signature. In global coordinates they were removed by the relation $ \e_i = \text{sgn}(\cos\s_i \cos\bar \s_i) $ which implies the fact that there is no distinction between the diamonds that we had in planar coordinates. 

Now, in planar coordinates, we analyze the singularity structure of the scalar leaf amplitude \eqref{lasc} in the cross ratio space defined by,
\be\label{cross_ratios}
z = \f{z_{12} z_{34}}{ z_{13} z_{24}}, \ \bar z = \f{\bar z_{12}\bar z_{34}}{\bar z_{13} \bar z_{24}}. 
\ee
 We first concentrate on the leaf amplitude in the timelike region. Since $ \mathcal{L}^s_4(z_i,\bar z_i) $ differs from $ \mathcal{C}_4 (z_i, \bar z_i) $ just by a constant we will concentrate on $ \mathcal{C}_4 (z_i, \bar z_i) $ only. So far we have kept $ \bar h_j $'s in \eqref{sa_pc1} arbitrary. We now focus on the scalar case where $ \bar{h}_j = \f{1+i\l_j}{2} $. For the sake of simplicity, we take the imaginary parts of the conformal dimensions of all the four external particles to be the same: $ \l_1=\l_2=\l_3=\l_4=\l $. 
Let us restore the first four arguments of the $ H $-function. With all $ \l_j = \l $ it is given by (see \ref{appB1}), 
 \be
 H(1+i\l,1+i\l,1+2i\l,2+2i\l; u_{\pm}, v_{\pm}) 
 \ee

For all $ \l_j = \l $ the delta function involving the total conformal weights becomes $ \d(\b)=\f{1}{4}\d(\l) $. On the support of this delta function $ H $-function becomes $ H(1,1,1,2; u, v)  $. It has been shown \cite{Dolan:2000uw} that for these specific arguments, $ H $-function can be written in terms of $ \log $ and $ \text{Li}_2 $ functions (See appendix \ref{H_func} for details). Then after properly taking care of the monodromies of these functions, it was shown \cite{Jain:2023fxc,Gary:2009ae,Alday:2024yyj} that $ H(1,1,1,2; u, v)  $ has simple pole singularity as the two cross ratios $ z $ and $ \bar z $ approach each other. We will now show this in detail. For concreteness we take 
\be \label{in_out}
\e_1 = \e_2=-1, \ \e_3=\e_4=+1.
\ee 
Also using conformal symmetry we set 
\be \label{conf_symm_coord}
 z_1,\bar z_1 \to \infty,  z_2=\bar{ z}_2 =1, \ z_3=z, \bar z_3=\bar z, z_4 =\bar z_4=0.
 \ee
Then, we define the following function,
\be\label{sleaf2}
\begin{split}
\mathcal{S}_4(z,\bar z) &= \lim_{z_1,\bar z_1 \to \infty} z_1^{2h_1} \bar z_1^{2\bar h_1} \mathcal{C}_4(z_i,\bar z_i), \ \ h_j = \bar h_j = \f{1+i\l}{2}, \forall j \\
&= \f{i\pi}{2} \G\(1+2i\l\) e^{2\pi\l} H(1+i\l,1+i\l,1+2i\l,2+2i\l; u_+, v_+) \\
& - \f{i\pi}{2} \G\(1+2i\l\) e^{-2\pi\l} H(1+i\l,1+i\l,1+2i\l,2+2i\l; u_-, v_-)
\end{split}
\ee
and, $ u_{\pm}, \, v_{\pm} $ become,
\be \label{sim_uv}
u_{\pm} = z \bar z \pm (1+z \bar z)i\e, \ v_{\pm} = (1-z)(1-\bar z) \pm \{(1-z)(1-\bar z) - 1\} i\e.
\ee
 The expressions of $ u_{\pm} $ and $ v_{\pm} $ in the above equation have been obtained using \eqref{in_out},  \eqref{conf_symm_coord} in \eqref{gen_uv}, then expanding the latter around $ \e=0 $ and keeping terms upto $ \mathcal{O}(\e) $. Instead of analysing the singularity structure of $ \mathcal{C}_4(z_i,\bar z_i) $ as $ z \to \bar z $, we will work with \eqref{sleaf2}. Since, $ \mathcal{C}_4(z_i,\bar z_i) $ is obtained from $ \mathcal{S}_4(z,\bar z) $ by multiplying a conformally covariant prefactor which does not contain any singularity as $ z \to \bar z $, working with $ \mathcal{S}_4(z,\bar z) $ is as good as $ \mathcal{C}_4(z_i,\bar z_i) $. 
 
Let us first start with the explicit expression of the $ H $-function given by,
\be\label{Dbar1111}
\begin{gathered}
H(1+i\l,1+i\l,1+2i\l,2+2i\l;u_{\pm},v_{\pm})= \f{1}{1-x_{\pm}-y_{\pm}}\[  \log\{x_{\pm}(1-y_{\pm})\}\log\(\f{y_{\pm}}{1-x_{\pm}}\) \right.\\
 \left. + 2 \{ Li_2(1-y_{\pm}) - Li_2(x_{\pm})\}\] + \mathcal{O}(\l)
\end{gathered}
\ee
where we have defined $ u_{\pm}=x_{\pm}(1-y_{\pm}), \ v_{\pm}=y_{\pm}(1-x_{\pm}) $. We have reviewed the derivation of \eqref{Dbar1111} in appendix \ref{H_func} for the sake of completeness of the paper. As explained in \ref{H_func}, we can identify $ x_{\pm}=z\mp i\e $ and $ 1-y_{\pm}=\bar z \pm i\e $ provided that $ \bar z > z >1 $. Hence, we assume $ \bar z > z >1 $ and write $ u_{\pm} $ and $ v_{\pm} $ as 
\be \label{upm_vpm}
u_{\pm} = (z\mp i\e)(\bar z \pm i\e), \ v_{\pm} = (1-z\pm i\e)(1-\bar z \mp i\e).
\ee

Then on the support of the delta function $ \d(\l) $ our $ H $-functions in \eqref{sleaf2} become
\be \label{H-func-1}
\begin{gathered}
\d(\l) H(1+i\l,1+i\l,1+2i\l,2+2i\l; u_{\pm}, v_{\pm}) \\
 = \f{\d(\l)}{\bar z - z \pm i\e}\[\{\log(z \mp i\e) + \log(\bar z \pm i\e)\} \{\log(1-\bar z \mp i\e) - \log(1-z \pm i\e) \} \right.\\
\left. - 2 \text{Li}_2(z \mp i\e) + 2 \text{Li}_2(\bar z \pm i\e) \]
\end{gathered} 
\ee

\begin{figure} [h]
\begin{center}
\includegraphics[height=8cm,width=10cm]{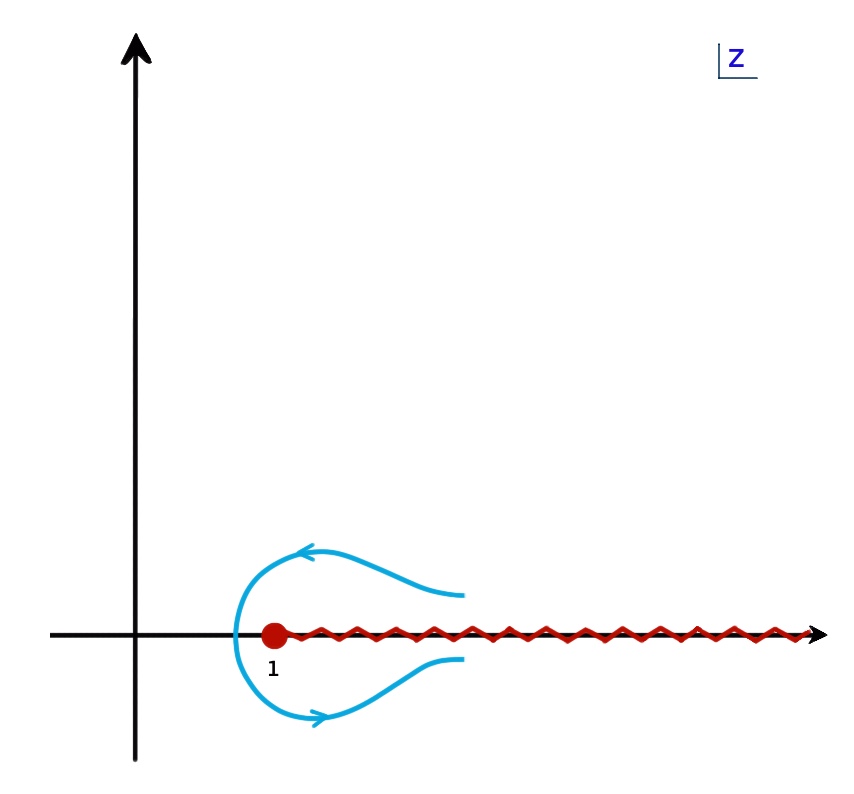}
\caption{Path traversed by $ z $ for timelike leaf amplitude.}
\label{fig_ana_cont}
\end{center}
\end{figure}

$ \log(1-x), \text{Li}_2(x) $ and $  \log x $ have nontrivial monodromies at branch point singularities $ 1 $ and $ \infty $. We  use the following branch cut prescription in the complex $ z, \bar z $ plane inspired by \citep{Gary:2009ae}. We place a branch cut from $ 1 $ to $ \infty $ on the positive real axis in the $ z $ plane and assume that $ z $ moves around the branch cut counterclockwise (see Fig \ref{fig_ana_cont}), whereas there is no branch cut in the $ \bar z $ plane. With this prescription we can take $ \e \to 0 $ limit in all the $ \log $ and $ \text{Li}_2 $ functions involving $ \bar z $ without encountering any discontinuities. On the other hand for $ z $ we have,
\be \label{mondr}
\begin{split}
 & \log(z+i\e) - \log(z - i\e) = -2\pi i, \ \log(1-z+i\e) - \log(1-z - i\e) = 2\pi i, \\
 & \hspace{4cm} \ \text{Li}_2( z + i\e) - \text{Li}_2( z - i\e) = 2\pi i \, \log( z ) 
\end{split}
\ee


Using the above discontinuities in the first $ H $-function of \eqref{sleaf2} we get 
\be \label{leaf_fin1}
\begin{split}
\d(\l)\mathcal{S}_4(z,\bar z) &= \f{i\pi}{2} \f{\d(\l)}{\bar z - z + i\e}\[4\pi^2 + 2\pi i \ln\left(\f{z}{\bar z}\f{1-\bar z}{1-z} \right) \], \ z, \bar z>1.
\end{split}
\ee

Thus we see that, on the support of the delta function arising from bulk scale invariance timelike leaf amplitude shows a simple pole singularity as $ z \to \bar z $. Spacelike leaf amplitude also has same singularity as we show in the next subsection and then we discuss how to get the celestial amplitude by adding the two leaf amplitudes.

\subsubsection{Recovering celestial amplitude from leaf amplitudes}
\label{recov}

In the previous section we have computed the leaf amplitude in the timelike region. After calculating the monodromies we have shown that it has a simple pole singularity at $ z = \bar z $. In the same way we can calculate the singularity structure of leaf amplitude in the spacelike region. The leaf amplitude in the spacelike region for arbitrary conformal weights is given in global coordinates in appendix \ref{appndxlc} (see \eqref{sa_pc2}). We now follow the same procedure as in the previous subsection. We first go to the spacelike leaf amplitude in planar coordinates $ \overline{\mathcal{C}_{4}}(z_i, \bar z_i) $ from the spacelike leaf amplitude in global coordinates  $  \mathcal{C}_{4}(\s_i, -\bar \s_i) $. Exploiting the conformal symmetry we then define a function $ \overline{\mathcal{S}_4}(z, \bar z)  $ like the one \eqref{sleaf2} we defined for the timelike leaf amplitude. We write the final expression,
\be\label{sls4}
\begin{gathered}
\overline{\mathcal{S}_4}(z, \bar z) = -\f{i\pi}{2}\Gamma(1+2i\l)H(1+i\l,1+i\l,1+2i\l,2+2i\l;u_+,v_+) \\
+ \f{i\pi}{2}\Gamma(1+2i\l)H(1+i\l,1+i\l,1+2i\l,2+2i\l;u_-,v_-) 
\end{gathered}
\ee


Now, we choose the paths traversed by $ z, \bar z $ to be exactly opposite as the time-like leaf amplitude, i.e., $ \bar z $ moves clockwise around the branch cut placed from $ 1 $ to $ \infty $ in the complex $ \bar z $ plane and there is no branch cut in the $ z $ plane. Then using the monodromies of the $ \log $ and $ \text{Li}_2 $ functions involving $ \bar z $, in the second $ H $-function of \eqref{sls4}, and expanding around $ \l=0 $ we get,

\be \label{leaf_fin2}
\begin{split}
\d(\l)\overline{\mathcal{S}_4}(z, \bar z) = \f{i\pi}{2} \f{\d(\l)}{\bar z - z - i\e}\[-4\pi^2 - 2\pi i \ln\(\f{z}{\bar z}\f{1-\bar z }{1-z} \) \], \ z,\bar z>1.
\end{split}
\ee

Celestial amplitude \eqref{cel_amp} is obtained by adding the two leaf amplitudes and then multiplying by $ \f{\d(\b)}{(2\pi)^3} $. Since we are working with the conformally invariant functions $ \mathcal{S}_4(z, \bar z) $ and $ \overline{\mathcal{S}_4}(z, \bar z) $, by adding them we will get the conformally invariant part of the celestial amplitude. The full celestial amplitude is then obtained by multiplying the conformally invariant part of the celestial amplitude with a conformally covariant prefactor which is fixed by conformal symmetries. Hence, from \eqref{leaf_fin1} and \eqref{leaf_fin2}, we get the conformally invariant part of the celestial amplitude as given by,
\be \label{cel_final}
\begin{gathered}
\widetilde{\mathcal{M}}_4(z, \bar z) = (-(2\pi)^4 \tilde{\l})\f{\d(\l)}{16} \Theta(z-1)\[\f{i}{\bar z - z + i\e} - \f{i}{\bar z - z - i\e} \] \\
= (-(2\pi)^4 \tilde{\l})\d(\l)\f{\pi}{8} \Theta(z-1)\d( z - \bar z )
\end{gathered}
\ee
where 
\be
\begin{split}
\Theta(z-1) &= 1, \ z>1\\
&=0, \ \text{otherwise} 
\end{split}
\ee

\eqref{cel_final} exactly matches with the conformally invariant part of the 4-point tree level scalar celestial amplitude derived in the appendix \ref{scal_cel_amp} (see equation \eqref{scal_mells}). Conformally covariant prefactor can be restored easily using the conformal symmetry. We want to emphasize that there are other prescriptions to calculate the monodromies of the functions appearing in leaf amplitudes. But not all of them will give us the required delta function of the cross ratios.

\subsection{MHV gluon scattering}
\label{mhv_gluon_scatt}

We now study the behavior of the MHV gluon leaf amplitudes in the timelike region as the two cross ratios $ z $ and $ \bar z $ approach each other. The color ordered 4-point MHV gluon amplitude is given by,
\be
\begin{gathered}
A_4\(1^{-}, 2^{-}, 3^{+}, 4^{+}\) = \f{\left<12\right>^3}{\left<23\right>\left<34\right>\left<41\right>} \d^{(4)}(p_1+p_2+p_3+p_4)
\end{gathered} 
\ee

To write the corresponding leaf amplitudes, we will first go to the global coordinates. In global coordinates the spinor brackets can be written as,
\be
 \left< ij \right> = \sqrt{\om_i \om_j} \sin\s_{ij} = \sqrt{\om_i \om_j} s_{ij}
\ee

From \eqref{msamp} we then see that,
\be \label{tlga}
A_4\(\left\{\e_i\om_i q_i, \vartheta_i \right\}\)  = \f{\om_1\om_2}{\om_3\om_4} \f{s_{12}^3}{s_{23}s_{34}s_{41}}
\ee

After replacing \eqref{tlga} in the first equation of \eqref{la1} we can write the timelike $ 4 $-point leaf amplitude for MHV gluon scattering as,

\be \label{oreq}
\begin{gathered}
\mathcal{L}_4^g(\s_i,\bar \s_i) = \f{s_{12}^3}{s_{23}s_{34}s_{41}} \mathcal{C}_4(\s_i,\bar \s_i)
\end{gathered} 
\ee

where $ \mathcal{C}_4(\s_i,\bar \s_i) $ is given by \eqref{main_res} with
\be \label{gl_wghts}
\begin{split}
\bar h_1 = 1 + \f{i\l_1}{2}, \ \bar h_2 = 1+ \f{i\l_2}{2}, \ \bar h_3 = \f{i\l_3}{2}, \ \bar h_4 = \f{i\l_4}{2}.
\end{split} 
\ee

Once again, we follow the same procedure as discussed in the scalar case in previous subsections. We first go to the planar coordinates using \eqref{plan_coord_tranf} and multiplying by the appropriate Jacobian factor. We also take $ \l_1=\l_2=\l_3=\l_4=\l $ and use \eqref{in_out}. Then we define the following function from the leaf amplitude in planar coordinates $ \mathcal{L}_4^g(z_i,\bar z_i) $ in the configuration \eqref{conf_symm_coord}: 
\be
\begin{gathered}\label{ga_pc1}
\mathcal{G}_4(z,\bar z) = \lim_{z_1, \bar z_1 \to \infty} z_1^{2h_1} \bar z_{1}^{2\bar h_1} \mathcal{L}^g_4(z_i,\bar z_i)\\
=\f{i\pi}{2} \G\(1 + 2i\l\) e^{2\pi \l}\f{z \bar z^2}{z-1} H(2 + i\l, 2 + i\l, 3+2i\l, 4+2i\l; u_+,v_+)\\
- \f{i\pi}{2}  \G\(1 + 2i\l\) e^{-2\pi \l}\f{z \bar z^2}{z-1}  H(2 + i\l, 2 + i\l, 3+2i\l, 4+2i\l; u_-,v_-)
\end{gathered} 
\ee

where $ u_{\pm}, \ v_{\pm} $ are given by \eqref{upm_vpm}. $ H $-functions appeared in \eqref{ga_pc1} satisfy the following two identities \cite{Dolan:2000uw, Dolan:2000ut},
\begin{eqnarray}
\label{fhi}
H(\a,\b,\g,\d;u,v) &=& H(\a,\b,\a+\b-\d+1,\a+\b-\g+1;v,u) \\
\nonumber (\d-\a-\b) \, H(\a,\b,\g,\d;u,v) &=& H(\a,\b,\g,\d+1;u,v) - v \, H(\a+1,\b+1,\g,\d+1;u,v) \\
\label{shi}
\end{eqnarray}

First using the identity \eqref{fhi}, then \eqref{shi} and finally \eqref{fhi} again, we can write
\be\label{hfunc_gluon}
\begin{gathered}
 H\(2 + i\l, 2 + i\l, 3+2i\l, 4+2i\l; u,v \) =  \f{1}{u} \[H(1+i\l,1+i\l,1+2i\l,2+2i\l;u,v) \right. \\
 \left. + (1+2i\l)H(1+i\l,1+i\l,2+2i\l,2+2i\l;u,v)\]
 \end{gathered}
\ee

Using \eqref{hfunc_gluon}, equation \eqref{ga_pc1} can be rewritten as
\be 
\mathcal{G}_{4}(z,\bar z) = \mathcal{G}_{4}^{\text{sing}}(z,\bar z) + \mathcal{G}_{4}^{\text{reg}}(z,\bar z)
\ee

where
\be \label{gluon_sing_reg}
\begin{split}
\mathcal{G}_{4}^{\text{sing}}(z,\bar z) &= \f{i\pi}{2} \G\(1 + 2i\l\) e^{2\pi \l}\f{ \bar z}{z-1} H(1 + i\l, 1 + i\l, 1+2i\l, 2+2i\l; u_+,v_+)\\
&- \f{i\pi}{2}  \G\(1 + 2i\l\) e^{-2\pi \l}\f{ \bar z}{z-1}  H(1 + i\l, 1 + i\l, 1+2i\l, 2+2i\l; u_-,v_-)\\
\mathcal{G}_{4}^{\text{reg}}(z,\bar z) &= \f{i\pi}{2} \G\(2 + 2i\l\) e^{2\pi \l}\f{ \bar z}{z-1} H(1 + i\l, 1 + i\l, 2+2i\l, 2+2i\l; u_+,v_+)\\
&- \f{i\pi}{2}  \G\(2 + 2i\l\) e^{-2\pi \l}\f{ \bar z}{z-1}  H(1 + i\l, 1 + i\l, 2+2i\l, 2+2i\l; u_-,v_-)
\end{split} 
\ee

The $ H $-function in the first equation of  \eqref{gluon_sing_reg} is the one that appeared in the scalar case and we have showed that they contain a simple pole singularity at $ z = \bar z $. As discussed in \cite{Dolan:2000uw} (equation C.11), the $ H $-function in the second equation of  \eqref{gluon_sing_reg} does not have any singularity at $ z = \bar z $. Thus using the same prescription for the paths of $ z, \bar z $ as described in the paragraph above  \eqref{mondr}, we can show that for $ \bar z > z>1 $,
\be \label{gluon_tl}
\d(\l) \mathcal{G}_{4}(z,\bar z) = \f{i\pi}{2} \f{\bar z}{z-1} \f{\d(\l)}{\bar z - z + i\e}\[4\pi^2 + 2\pi i \log\(\f{z}{\bar z}\f{1-\bar z }{1-z} \) \]  + \d(\l)\text{Reg}^t.
\ee
where $ \text{Reg}^t $ is the regular terms as $ z \to \bar{ z} $. Similarly, for $ \bar z> z >1 $ the spacelike leaf amplitude is given by
\be \label{gluon_sl}
\d(\l) \overline{\mathcal{G}_{4}}(z,\bar z) = \f{i\pi}{2} \f{\bar z}{z-1} \f{\d(\l)}{\bar z - z + i\e}\[-4\pi^2 - 2\pi i \log\(\f{z}{\bar z}\f{1-\bar z }{1-z} \) \]  + \d(\l) \text{Reg}^s.
\ee

By adding \eqref{gluon_tl} and \eqref{gluon_sl} and multiplying $ \f{1}{(2\pi)^3} \f{1}{4}$, we get
\be \label{gluon_cel_fl}
\d(\l)\f{\pi}{8} \Theta(z-1)\f{ z}{z-1} \d( z - \bar z ) + \d(\l) \( \text{Regular terms as} \ z \to \bar z \)
\ee

The distributional part in the above equation exactly matches with conformally invariant part of the $ 4 $-point celestial MHV gluon amplitudes. This can be seen by applying our configuration \eqref{in_out},\eqref{conf_symm_coord} to 4-point MHV gluon celestial amplitude computed in \cite{Pasterski:2017ylz} and extracting the conformally invariant part which only depends on the cross ratios $ z $ and $\bar z$.  The regular terms are expected to vanish from the final celestial amplitude, i.e. form \eqref{gluon_cel_fl}. We hope to address this in future.

\section{BG equations for MHV gluon leaf amplitudes}
\label{BG_eq}

MHV gluon leaf amplitudes can be shown to satisfy BG equations first derived in \cite{Banerjee:2020vnt} \footnote{ See \cite{Hu:2021lrx} for momentum space origin of these differential equations.} by considering the subleading ($ \mathcal{O}(1) $) terms in the OPE between two positive helicity outgoing gluon operators and soft gluon theorems. In this section we derive the BG equations for the timelike leaf amplitudes in the same way. We will work with the $ 4 $-point amplitude with $ \e_1=\e_2=-1, \e_3=\e_4=+1 $. We first write the $ 4 $-point leaf amplitude in the planar coordinates in a form that is appropriate for OPE decomposition.  We also restore the color indices. So we start with the full $ 4 $-point tree level MHV gluon amplitude in momentum space as given by:
\be \label{full4point}
\begin{gathered}
A_4(1^{-,a_1},2^{-,a_2},3^{+,a_3},4^{+,a_4}) = g_{YM}^2 \left\{ A_4[1^-2^-3^+4^+] \text{tr}\(T^{a_1}T^{a_2}T^{a_3} T^{a_4}\) + \text{perm}(234) \right\} \\
\hspace{4cm} \times \d^{(4)}\(\sum_{i=1}^4 p_i\)
\end{gathered} 
\ee
where $ g_{YM} $ is the coupling constant  and $  A_4[i^{\vartheta_i}j^{\vartheta_j}k^{\vartheta_k}l^{\vartheta_l}] $ are color ordered partial MHV amplitudes given by,
\be\label{partial4point}
 A_4[i^-j^+k^-l^+]=\f{\left< ik\right>^4}{\left< ij\right> \left< jk\right> \left< kl\right> \left< li\right>}
\ee
 Substituting \eqref{partial4point} in \eqref{full4point} we get \footnote{Throughout this section we will use planar coordinates and follow the same  parametrization of the spinor brackets as given in \cite{Melton:2023bjw}.},
\be
\begin{gathered}
A_4(1^{-,a_1},2^{-,a_2},3^{+,a_3},4^{+,a_4}) = -g_{YM}^2 \f{\om_1 \om_2}{\om_3 \om_4} \f{z_{12}^3}{z_{23}z_{34}z_{41}}\[ f^{a_1 a_2 x}f^{x a_3 a_4} \right.\\
\left. \hspace{4cm} - \f{z_{12}z_{34}}{z_{13}z_{24}}f^{a_1 a_3 x} f^{x a_2 a_4} \] \d^{(4)}\(\sum_{i=1}^4 p_i\)
\end{gathered} 
\ee
In planar coordinates the $ 4 $-point leaf amplitude in the timelike region corresponding to the above amplitude is given by,
\be  \label{4point_AO}
\begin{gathered}
\mathcal{L}_4^g(1^{-,a_1}_{\D_1},2^{-,a_2}_{\D_2},3^{+,a_3}_{\D_3},4^{+,a_4}_{\D_4}) = -g_{YM}^2 \f{z_{12}^3}{z_{23}z_{34}z_{41}} \[ f^{a_1 a_2 x}f^{x a_3 a_4} - \f{z_{12}z_{34}}{z_{13}z_{24}}f^{a_1 a_3 x} f^{x a_2 a_4} \] \\
\times \int_{\hat{x}_+^2=-1}  d^3 \hat{x}_+  \left( \prod_{j=1}^2 \int_0^\infty d\om_j \, \om_j^{\D_j} \right)\left( \prod_{j=3}^4 \int_0^\infty d\om_j \, \om_j^{\D_j-2} \right) e^{ \sum_{k=1}^4 (i\e_k \om_k q_k \cdot \hat{x}_+ -\e \om_k)}
\end{gathered}
\ee

where $ q^\mu_k $ in terms of $ (z_k,\bar z_k) $ is given by \eqref{pmu} and we have used a different notation for the leaf amplitudes which is convenient for OPE factorization. $ i^{\pm}_{\D_i} $ inside $ \mathcal{L}_4^g $ means $ i $-th gluon with helicity $ \vartheta_i = \pm 1 $ and conformal dimension $ \D_i $. In terms of the scalar conformal primary wavefunction
\be\label{scal_conf_prim}
\Phi_{\D_i,\e_i}(\hat{x}, q_i) = \int_{0}^\infty d\om_i \, \om_i^{\D_i-1} e^{i \e_i \om_i q_i \cdot \hat{x} - \e \om_i} 
\ee

we can write the leaf amplitude \eqref{4point_AO} as,
\be\label{Lphi}
\begin{gathered}
\mathcal{L}_4^g(1^{-,a_1}_{\D_1},2^{-,a_2}_{\D_2},3^{+,a_3}_{\D_3},4^{+,a_4}_{\D_4}) = -g_{YM}^2 \f{z_{12}^3}{z_{23}z_{34}z_{41}} \[ f^{a_1 a_2 x}f^{x a_3 a_4} - \f{z_{12}z_{34}}{z_{13}z_{24}}f^{a_1 a_3 x} f^{x a_2 a_4} \] \\
\times \int_{\hat{x}_+^2=-1}  d^3 \hat{x}_+ \(\prod_{j=1}^2 \Phi_{\D_j+1,-}(\hat{x}_+,q_j) \) \(\prod_{j=3}^4 \Phi_{\D_j-1,+}(\hat{x}_+,q_j) \)
\end{gathered} 
\ee

We want to compute the OPE between the positive helicity outgoing \footnote{Here we abuse the notation and call the conformal gluon operator corresponding to positive frequency solutions in the bulk as outgoing operator.} conformal gluon operators inserted at points $ (z_3, \bar z_3) $ and $ (z_4, \bar z_4) $ on the celestial torus. Different modes of the leading and subleading soft gluon symmetry algebra will appear in the subleading order ($ \mathcal{O}(1) $) of the OPE. Hence, following \cite{Guevara:2021abz,Strominger:2021lvk,Strominger:2021mtt,Fan:2019emx,Donnay:2018neh,Pate:2019mfs,Nandan:2019jas,Adamo:2019ipt}, we define the leading and subleading conformally soft gluon operators as,
\be
R^{k,a} (z,\bar z) := \lim_{\D \to k}(\D-k) \mathcal{O}^{+,a}_{\D}(z, \bar z), \ k=1,0.
\ee
where $ \mathcal{O}^{+,a}_{\D}(z, \bar z) $ denote a positive helicity outgoing gluon conformal primary operator of dimension $ \D $ at the point $ (z, \bar z) $ on the celestial torus. In fact one can define a tower of conformally soft gluon operators and the corresponding conserved currents follow a symmetry algebra known as $ S $-algebra \cite{Strominger:2021mtt}. It was shown in \cite{Melton:2024jyq} that MHV gluon leaf amplitudes respect this symmetry algebra. However, for our purpose we will restrict to leading and subleading soft gluon symmetry. The soft current $ R^{1,a}_0(z) $ is a Kac-Moody current \cite{Strominger:2014ass,He:2014cra,Kapec:2015ena,Kapec:2014zla,Campiglia:2015qka,Nande:2017dba,He:2020ifr}.

The explicit expressions of the actions of the modes $ R^{k,a}_{\a,n} $ of the operators $ R^{k,a}(z,\bar z) $ ($ k=1,0 $) on the amplitudes \cite{Banerjee:2020vnt} can be found out from the known soft theorems. In particular, for our purpose the leading soft gluon mode that will play an important role, is given by $ R^{1,a}_{-1,0} $. Its action on the $ 3 $-point leaf amplitudes
\be\label{3pointleaf}
\begin{gathered}
\mathcal{L}_3^g\(1_{\D_1}^{-,a_1}, 2_{\D_2}^{-,a_2}, 4^{+,a_4}_{\D_3+\D_4-1}\) = -2i g_{YM} f^{a_1 a_2 a_4} \f{z_{12}^3}{z_{24}z_{41}} \int d^3 \hat{x}_+ \left( \prod_{j=1}^2 \Phi_{\D_j+1,-}(\hat{x}_+,q_j) \right) \\
 \hspace{4cm} \times \Phi_{\D_3+\D_4-2,+}\(\hat{x}_+,q_4\)
\end{gathered} 
\ee
can be determined from the leading positive helicity soft gluon theorem and it is given by \footnote{Equation \eqref{lead_action} can be obtained by applying equation (3.9) of \cite{Banerjee:2020vnt} to the $ 3 $-point leaf amplitudes and by identifying $ \mathcal{J}^a_{-1} = \mathcal{R}^{1,a}_{-1,0} $.},
\be\label{lead_action}
\begin{gathered}
\mathcal{R}^{1,a_3}_{-1,0}\mathcal{L}^g_3\(1_{\D_1}^{-,a_1}, 2_{\D_2}^{-,a_2}, 4_{\D_3+\D_4-1}^{+,a_4}\) = -2g_{YM}\( \f{f^{a_1 a_3 x}f^{x a_2 a_4}}{z_{14}} + \f{f^{x a_2 a_3} f^{x a_4 a_1}}{z_{24}}\) \\
\times \f{z_{12}^3}{z_{24}z_{41}} \int d^3 \hat{x}_+  \left( \prod_{j=1}^2 \Phi_{\D_j+1,\e_j}(\hat{x}_+,q_j) \right) \Phi_{\D_3+\D_4-2,+}\(\hat{x}_+,q_4\)
\end{gathered} 
\ee

Let us now move to the OPE factorization.

\subsection{OPE factorization}
We will extract the OPE by factorizing the four point gluon leaf amplitudes. Since we are interested in computing the OPE between $ 3 $rd and $ 4 $th gluon primaries, we will concentrate on the product of the following two scalar conformal primaries,
\be \label{Phi34}
\begin{gathered}
\Phi_{\D_3-1,+}\(\hat{x}_+,q_3\) \Phi_{\D_4-1,+}\(\hat{x}_+,q_4\) = \int_{0}^\infty d\om_3 \, \om_3^{\D_3-2} \int_{0}^\infty d\om_4 \, \om_4^{\D_4-2}e^{i(\om_3 q_3+\om_4 q_4)\cdot \hat{x}_+}
\end{gathered} 
\ee

Using the following parametrization:
\be
\om_3 = t\om, \ \om_4=(1-t)\om 
\ee

and expanding around $ z_{34}=0, \bar z_{34} = 0 $ we get from \eqref{Phi34},
\be
\begin{gathered}
\Phi_{\D_3-1,+}\(\hat{x}_+,q_3\) \Phi_{\D_4-1,+}\(\hat{x}_+,q_4\) = B(\D_3-1,\D_4-1) \Phi_{\D_3+\D_4-2,+}\(\hat{x}_+,q_4\)\\
+ z_{34} \, B(\D_3,\D_4-1) \f{\pa}{\pa z_4}\Phi_{\D_3+\D_4-2,+}\(\hat{x}_+,q_4\) + \cdots
\end{gathered} 
\ee

We use the above equation in \eqref{Lphi} and expanding around $ z_{34}=0, \bar z_{34} = 0 $ we get,
\be \label{4ptexpan}
\begin{gathered}
\mathcal{L}_4^g\(1_{\D_1}^{-,a_1}, 2_{\D_2}^{-,a_2}, 3_{\D_3}^{+,a_3}, 4_{\D_4}^{+,a_4}\) = -g_{YM}^2 \f{1}{z_{34}} B(\D_3-1,\D_4-1) f^{a_1 a_2 x}f^{x a_3 a_4} \\
\times \f{z_{12}^3}{z_{24}z_{41}} \int d^3 \hat{x}_+ \left( \prod_{j=1}^2 \Phi_{\D_j+1,\e_j}(\hat{x}_+,q_j) \right) \Phi_{\D_3+\D_4-2,+}\(\hat{x}_+,q_4\)\\
- g_{YM}^2 B(\D_3-1,\D_4-1)\( \f{f^{a_1 a_3 x}f^{x a_2 a_4}}{z_{14}} + \f{f^{x a_2 a_3} f^{x a_4 a_1}}{z_{24}}\) \\
\times \f{z_{12}^3}{z_{24}z_{41}} \int d^3 \hat{x}_+ \left( \prod_{j=1}^2 \Phi_{\D_j+1,\e_j}(\hat{x}_+,q_j) \right) \Phi_{\D_3+\D_4-2,+}\(\hat{x}_+,q_4\)\\
 -g_{YM}^2 f^{a_1 a_2 x}f^{x a_3 a_4} \, B(\D_3,\D_4-1)  \f{z_{12}^3}{z_{24}z_{41}} \\
\times \f{\pa}{\pa z_4} \int d^3 \hat{x}_+\left( \prod_{j=1}^2 \Phi_{\D_j+1,\e_j}(\hat{x}_+,q_j) \right)  \Phi_{\D_3+\D_4-2,+}\(\hat{x}_+,q_4\) + \cdots
\end{gathered}
\ee

We are now ready to write down the OPE. The leading order term was calculated in \cite{Melton:2024jyq} and is given by,

\be
\mathcal{O}^{+,a_3}_{\D_3}(z_3, \bar z_3) \mathcal{O}^{+,a_4}_{\D_4}(z_4, \bar z_4) =  - \f{g_{YM}}{2} \f{1}{z_{34}} B(\D_3-1,\D_4-1) if^{a_3 a_4 x} \mathcal{O}^{+,x}_{\D_3+\D_4-1}(z_4, \bar z_4)
\ee
This is the same as the leading order OPE between two positive helicity gluons obtained from celestial MHV amplitude \cite{Pate:2019lpp}. We now discuss the $ \mathcal{O}(1) $ term.

\subsubsection{Subleading terms: $ \mathcal{O}(1) $}

Using \eqref{lead_action} in \eqref{4ptexpan} at $ \mathcal{O}(1) $ we have,

\be \label{4ptO1}
\begin{gathered}
\mathcal{L}^g_4\(1_{\D_1}^{-,a_1}, 2_{\D_2}^{-,a_2}, 3_{\D_3}^{+,a_3}, 4_{\D_4}^{+,a_4}\)\big|_{\mathcal{O}(1)} = \f{g_{YM}}{2}\[ B(\D_3,\D_4-1)(-if^{x a_3 a_4}) \mathcal{L}_{-1}(4)\mathcal{L}^g_3\(1_{\D_1}^{-,a_1}, 2_{\D_2}^{-,a_2}, 4_{\D_3+\D_4-1}^{+,x}\) \right. \\
\left. + B(\D_3-1,\D_4-1)\mathcal{R}^{1,a_3}_{-1,0}\mathcal{L}^g_3\(1_{\D_1}^{-,a_1}, 2_{\D_2}^{-,a_2}, 4_{\D_3+\D_4-1}^{+,a_4}\)  - B(\D_3,\D_4-1)\mathcal{R}^{1,a_3}_{-1,0}\mathcal{L}^g_3\(1_{\D_1}^{-,a_1}, 2_{\D_2}^{-,a_2}, \right.\right. \\
\left.\left. 4_{\D_3+\D_4-1}^{+,a_4}\)
+ B(\D_3,\D_4-1)\mathcal{R}^{1,a_4}_{-1,0}\mathcal{M}^+_3\(1_{\D_1}^{-,a_1}, 2_{\D_2}^{-,a_2}, 4_{\D_3+\D_4-1}^{+,a_3}\) \]
\end{gathered}
\ee

After some simplification the above equation can be written in terms of the correlators as follows,
\be \label{4ptfinal}
\begin{gathered}
\left< \mathcal{O}^{-,a_1}_{\D_1,-}(z_1, \bar z_1) \mathcal{O}^{-,a_2}_{\D_2,-}(z_2, \bar z_2) \mathcal{O}^{+,a_3}_{\D_3,+}(z_3, \bar z_3) \mathcal{O}^{+,a_4}_{\D_4,+}(z_4, \bar z_4)\right>\big|_{\mathcal{O}(1)} \\
= \f{g_{YM}}{2} \left[ B(\D_3,\D_4-1)(-if^{x a_3 a_4}) \left< \mathcal{O}^{-,a_1}_{\D_1,-}(z_1, \bar z_1) \mathcal{O}^{-,a_2}_{\D_2,-}(z_2, \bar z_2) L_{-1}\mathcal{O}^{+,x}_{\D_3+\D_4-1,+}(z_4, \bar z_4)\right> \right. \\
\left. + B(\D_3-1,\D_4-1)\[ \f{\D_4-1}{\D_3+\D_4-2} \left< \mathcal{O}^{-,a_1}_{\D_1,-}(z_1, \bar z_1) \mathcal{O}^{-,a_2}_{\D_2,-}(z_2, \bar z_2) R^{1,a_3}_{-1,0} \mathcal{O}^{+,a_4}_{\D_3+\D_4-1,+}(z_4, \bar z_4)\right> \right.\right. \\
\left.\left. + \f{\D_3-1}{\D_3+\D_4-2}\left< \mathcal{O}^{-,a_1}_{\D_1,-}(z_1, \bar z_1) \mathcal{O}^{-,a_2}_{\D_2,-}(z_2, \bar z_2) R^{1,a_4}_{-1,0} \mathcal{O}^{+,a_3}_{\D_3+\D_4-1,+}(z_4, \bar z_4)\right> \] \]
\end{gathered}
\ee

Hence, at the level of OPE we have,
\be \label{O1OPE}
\begin{gathered}
 \mathcal{O}_{\D_3}^{+,a_3}(z_3,\bar z_3) \mathcal{O}_{\D_4}^{+,a_4}(z_4, \bar z_4)\big|_{\mathcal{O}(1)} = \f{g_{YM}}{2} \left[ B(\D_3,\D_4-1)(-if^{x a_3 a_4}) L_{-1} \mathcal{O}_{\D_3+\D_4-1}^{+,x}(z_4, \bar z_4 ) \right.\\
\left.+ B(\D_3-1,\D_4-1)\[ \f{\D_4-1}{\D_3+\D_4-2} R^{1,a_3}_{-1,0} \mathcal{O}_{\D_3+\D_4-1}^{+,a_4}(z_4, \bar z_4 ) \right.\right. \\
\left.\left. + \f{\D_3-1}{\D_3+\D_4-2} R^{1,a_4}_{-1,0}\mathcal{O}_{\D_3+\D_4-1}^{+,a_3}(z_4, \bar z_4 ) \] \right]
\end{gathered}
\ee

This OPE exactly matches with the OPE obtained from celestial MHV gluon amplitudes in \cite{Banerjee:2020vnt,Ebert:2020nqf}.

\subsection{BG equations}
 Now, we can take the subleading soft limit in \eqref{O1OPE} and get the following null equation \cite{Banerjee:2020vnt},
\be \label{null_eq}
\begin{gathered}
if^{abc}L_{-1} \mathcal{O}^{+,c}_{\D-1,+}(z, \bar z) + R^{0,a}_{-\f{1}{2},\f{1}{2}} \mathcal{O}^{+,b}_{\D,+}(z,\bar z) - R^{1,b}_{-1,0} \mathcal{O}^{+,a}_{\D-1,+}(z, \bar z) \\
+ (\D-1)R^{1,a}_{-1,0} \mathcal{O}^{+,b}_{\D-1,+}(z, \bar z) = 0
\end{gathered}
\ee
decoupling of which from the leaf amplitude will lead to BG equations \cite{Banerjee:2020vnt} for the latter. In a theory which respects leading and subleading positive helicity soft gluon theorems is guaranteed to have the above null state. Since, $ S $-algebra does not change for leaf amplitudes, it was expected that the leaf amplitudes will satisfy the BG equations, which are the differential equations corresponding to the null state \eqref{null_eq}.

\section*{Acknowledgement}

We are grateful to Shamik Banerjee for suggesting this problem to us. We also  thank Shamik Banerjee, Alok Laddha and Bobby Ezhuthachan for various helpful discussions and insightful comments on the draft.

\appendix

\section{Detailed calculation: 4-point scalar leaf amplitude}
\label{appndxlc}

In this section we perform the integral \eqref{gen_c4} in detail and calculate the 4-point scalar leaf amplitudes both in time-like and space-like regions. We will use the techniques described in \cite{Melton:2023bjw} and \cite{Penedones:2016voo}. The detailed notations are given in section \ref{revw}. In global coordinates the measure is given by, 
\be
d^3\hat{x}_+ = \sinh\rho\cosh\rho \, d\rho d\psi d\phi
\ee

Using \eqref{smallx} and \eqref{phat}, one can compute the following,
\be
\begin{gathered}
\hat{p}_k \cdot \hat{x}_+ = \cos(\phi-\phi_k)\sinh\rho - \cos(\psi-\psi_k)\cosh\rho \\
\end{gathered} 
\ee

Then the integral \eqref{gen_c4} becomes,
\be \label{eq:A3}
\begin{gathered}
\mathcal{C}_4(\s_i,\bar \s_i) = \int_{0}^\infty d\rho \, \sinh\rho \, \cosh\rho \int_0^{2\pi}d\psi \int_0^{2\pi} d\phi \( \prod_{k=1}^4 \int_0^\infty d\om_k \om_k^{2\bar h_k -1} e^{- \e \om_k} \)\\
\times e^{i x \cos\phi + i y \sin\phi} e^{i\bar x \cos\psi + i \bar y \sin\psi}
\end{gathered} 
\ee
where 
\be \label{eq:A4}
\begin{gathered}
x = \sinh\rho \sum_{k=1}^4 \om_k \cos\phi_k, \ y = \sinh\rho \sum_{k=1}^4 \om_k \sin\phi_k \\
\bar x = -\cosh\rho \sum_{k=1}^4 \om_k \cos\psi_k, \ \bar y = -\cosh\rho \sum_{k=1}^4 \om_k \sin\psi_k
\end{gathered}
\ee
Remember that $ \s_i, \bar \s_i $ are the global coordinates on celestial torus and they are related to $ \psi_i, \phi_i $ via \eqref{null_coord}. We can perform the $ \phi $ and $ \psi $ integrals in \eqref{eq:A3} to get,
\be \label{eq:A5}
\begin{gathered}
\mathcal{C}_4(\s_i,\bar \s_i) = 4\pi^2 \int_{0}^\infty d\rho \, \sinh\rho \, \cosh\rho \( \prod_{k=1}^4 \int_0^\infty d\om_k \om_k^{2\bar h_k -1} e^{- \e \om_k} \)\\
\times J_0\(\sqrt{x^2+y^2}\) J_0\(\sqrt{\bar x^2+\bar y^2}\)
\end{gathered} 
\ee
One can check using \eqref{eq:A4} that,
\be
\begin{split}
\sqrt{x^2+y^2} &= \sinh\rho \sqrt{\sum_{k=1}^4 \om_k^2 + 2\sum_{j<k}^4 \om_j\om_k \cos\phi_{jk}} =  \sinh\rho \, \Phi \\
\sqrt{\bar x^2+\bar y^2} &= \cosh\rho \sqrt{\sum_{k=1}^4 \om_k^2 + 2\sum_{j<k}^4 \om_j\om_k \cos\psi_{jk}} =  \cosh\rho \, \Psi
\end{split} 
\ee
where $ \phi_{ij}=\phi_i - \phi_j, \ \psi_{ij} = \psi_i - \psi_j $. Next we make a change of variables $ y = \sinh\rho $ in \eqref{eq:A5} to get,
\be \label{eq:A6}
\begin{gathered}
\mathcal{C}_4(\s_i,\bar \s_i) = 4\pi^2 \( \prod_{k=1}^4 \int_0^\infty d\om_k \om_k^{2\bar h_k -1} e^{- \e \om_k} \) \int_{0}^\infty dy y J_0\(y \Phi \) J_0\( \sqrt{1+y^2}\Psi\)
\end{gathered} 
\ee

Using the following representation of the Bessel function
\be
J_0\( \sqrt{1+y^2}\Psi\) = \f{1}{2\pi i} \int_{\d-i\infty}^{\d+i\infty} \f{d\xi}{\xi} e^{\xi - \f{(1+y^2)\Psi^2}{4\xi}}
\ee
and rescaling $ \xi \to \f{\xi \Psi^2}{4} $ in \eqref{eq:A6} gives,
\be \label{eq:A8}
\begin{gathered}
\mathcal{C}_4(\s_i,\bar \s_i) = -2\pi i \( \prod_{k=1}^4 \int_0^\infty d\om_k \om_k^{2\bar h_k -1} e^{- \e \om_k} \) \int_{0}^\infty dy y J_0\(y \Phi \) \\
\times \int_{\d-i\infty}^{\d+i\infty} \f{d\xi}{\xi} e^{\f{\xi \Psi^2}{4} - \f{(1+y^2)}{\xi}}
\end{gathered} 
\ee

Then one can easily perform the $ y $-integral in the above equation and get,
\be
\begin{gathered}
\mathcal{C}_4(\s_i,\bar \s_i) = -\pi i \( \prod_{k=1}^4 \int_0^\infty d\om_k \om_k^{2\bar h_k -1} e^{- \e \om_k} \)  \int_{\d-i\infty}^{\d+i\infty} d\xi \, e^{-\f{1}{\xi} + \f{\xi}{4} (\Psi^2 - \Phi^2) }
\end{gathered} 
\ee
with
\be
\Psi^2 - \Phi^2 = -4\sum_{j<k}^4 \om_j \om_k s_{jk} \bar s_{jk}
\ee

Hence, the scalar leaf amplitude can be written as,
\be
\begin{gathered}
\mathcal{C}_4(\s_i,\bar \s_i) = -\pi i \int_{\d-i\infty}^{\d+i\infty} d\xi \, e^{-\f{1}{\xi}} I_4
\end{gathered} 
\ee
where 
\be 
I_4 =  \( \prod_{k=1}^4 \int_0^\infty d\om_k \om_k^{2\bar h_k -1} e^{- \e \om_k} \)   e^{-\xi\sum_{j<k}^4 \om_j \om_k s_{jk} \bar s_{jk}}
\ee
We now compute the above integral. There are six terms in the summation, namely,
\be
\begin{gathered}
\om_1 \om_2 s_{12} \bar s_{12} + \om_1 \om_3 s_{13} \bar s_{13} + \om_1 \om_4 s_{14} \bar s_{14} + \om_2 \om_3 s_{23} \bar s_{23} + \om_2 \om_4 s_{24} \bar s_{24} + \om_3 \om_4 s_{34} \bar s_{34}
\end{gathered} 
\ee

We use the Mellin-Barnes representation for the following two exponentials,
\be
\begin{gathered}
e^{-\xi \om_1 \om_2 s_{12} \bar s_{12}}e^{-\xi \om_1 \om_4 s_{14} \bar s_{14}} = \int_{c-i\infty}^{c+i\infty}\f{ds}{2\pi i} \G(s)\(\xi \om_1 \om_4 s_{14} \bar s_{14}\)^{-s} \\
\times \int_{c'-i\infty}^{c'+i\infty}\f{dr}{2\pi i} \G(r)(\xi \om_1 \om_2 s_{12} \bar s_{12})^{-r}
\end{gathered} 
\ee
 Let's first write $ I_4 $ as,
\be 
\begin{gathered}
I_4 =  \int_{c-i\infty}^{c+i\infty}\f{ds}{2\pi i} \G(s)\(\xi s_{14} \bar s_{14}\)^{-s}\int_{c'-i\infty}^{c'+i\infty}\f{dr}{2\pi i} \G(r)(\xi s_{12} \bar s_{12})^{-r} \( \prod_{k=2}^4 \int_0^\infty d\om_k \om_k^{2\bar h_k -1} e^{- \e \om_k} \)\\
\times \om_4^{-s} \om_2^{-r}  e^{-\xi(\om_2 \om_3 s_{23} \bar s_{23} + \om_2 \om_4 s_{24} \bar s_{24} + \om_3 \om_4 s_{34} \bar s_{34})} \int_0^\infty d\om_1 \om_1^{2\bar h_1 -r-s-1} e^{- \om_1 (\xi \om_3 s_{13} \bar s_{13} + \e)}
\end{gathered}
\ee

Then performing the $ \om_1 $ integral we get,
\be 
\begin{gathered}
I_4 =  \int_{c-i\infty}^{c+i\infty}\f{ds}{2\pi i} \G(s)\(\xi s_{14} \bar s_{14}\)^{-s}\int_{c'-i\infty}^{c'+i\infty}\f{dr}{2\pi i} \G(r)(\xi s_{12} \bar s_{12})^{-r} \f{\G(2\bar h_1 - r - s)}{(\xi s_{13} \bar s_{13} + \e)^{2\bar h_1 - r - s}}\\
\times \( \prod_{k=2}^4 \int_0^\infty d\om_k \om_k^{2\bar h_k -1} e^{- \e \om_k} \)\om_4^{-s} \om_2^{-r} \om_3^{r+s-2\bar h_1} e^{-\xi(\om_2 \om_3 s_{23} \bar s_{23} + \om_2 \om_4 s_{24} \bar s_{24} + \om_3 \om_4 s_{34} \bar s_{34})}
\end{gathered}
\ee

Let us now make the change of variable $ \om_k = \f{\sqrt{ t_2 t_3 t_4}}{t_k} $ and perform the $ t $ integrals. Thus $ I_4 $ becomes,

\be 
\begin{gathered}
I_4 =   \f{1}{2} (\xi s_{13} \bar s_{13} + \e)^{-2\bar h_1}(\xi s_{34} \bar s_{34} + \e)^{\bar h_1 +\bar h_2 - \bar h_3 - \bar h_4} (\xi s_{24} \bar s_{24}+\e)^{-\bar h_1 - \bar h_2 + \bar h_3 - \bar h_4} \\
\times (\xi s_{23} \bar s_{23} + \e)^{\bar h_1 - \bar h_2 - \bar h_3 + \bar h_4} \mathcal{F}_4(\tilde{u},\tilde{v})
\end{gathered}
\ee

where
\be \label{Irs}
\begin{gathered}
\mathcal{F}_4(\tilde{u},\tilde{v}) =   \int_{c-i\infty}^{c+i\infty}\f{ds}{2\pi i} \int_{c'-i\infty}^{c'+i\infty}\f{dr}{2\pi i} \G(s)\G(r) \G(s-\bar h_1 + \bar h_2 + \bar h_3 - \bar h_4) \\
\times \G(r-\bar h_1 - \bar h_2 + \bar h_3 + \bar h_4) \G(2\bar h_1 - r - s) \G(-r-s+\bar h_1 + \bar h_2 - \bar h_3 + \bar h_4) \tilde{u}^{-r}\tilde{v}^{-s} 
\end{gathered}
\ee
and we have defined,
\be
\tilde{u} = \f{(\xi s_{12} \bar s_{12})(\xi s_{34} \bar s_{34} + \e)}{(\xi s_{13} \bar s_{13} + \e)(\xi s_{24} \bar s_{24} + \e)}, \tilde{v} = \f{\(\xi s_{14} \bar s_{14}\)(\xi s_{23} \bar s_{23}+\e)}{(\xi s_{13} \bar s_{13} + \e)(\xi s_{24} \bar s_{24} + \e)}
\ee

We now change $ r \to -r, \ s \to -s $ in the integral \eqref{Irs} and get,
\be \label{Irsm}
\begin{gathered}
\mathcal{F}_4(\tilde{u},\tilde{v}) =   \int_{-c-i\infty}^{-c+i\infty}\f{ds}{2\pi i} \int_{-c'-i\infty}^{-c'+i\infty}\f{dr}{2\pi i} \G(-s)\G(-r) \G(-s-\bar h_1 + \bar h_2 + \bar h_3 - \bar h_4) \\
\times \G(-r-\bar h_1 - \bar h_2 + \bar h_3 + \bar h_4) \G(2\bar h_1 + r + s) \G(r+s+\bar h_1 + \bar h_2 - \bar h_3 + \bar h_4) \tilde{u}^{r}\tilde{v}^{s} 
\end{gathered}
\ee
$ c,c'>0 $. The above integral is given by (B.9) of \cite{Dolan:2000uw} and can be written in terms of the $ H $-function defined in that paper. More specifically, we can write,
\be \label{H-func}
\begin{gathered}
\mathcal{F}_4(\tilde{u},\tilde{v}) = H\(2\bar h_1 , \bar h_1 + \bar h_2 - \bar h_3+\bar h_4, 2\bar h_1+ 2 \bar h_2-1, 2\bar h_1+ 2 \bar h_2;\tilde{u},\tilde{v}\) \\
\end{gathered}
\ee

We will ignore writing the first four arguments of the $ H $-function unless their explicit expressions are required. Since the $ \xi $-dependence in $ H $-function is only through $ \tilde{u},\tilde{v} $ (and hence becomes $ \xi $-independent by giving proper $ i\e $ factors), we can easily perform the $ \xi $ integral. Let us write the scalar leaf amplitudes $ \mathcal{C}_4(\s_i,\bar \s_i) $ in terms of the $ H $-function,
\be
\begin{gathered}
\mathcal{C}_4(\s_i,\bar \s_i) = -\f{i\pi}{2} \int_{\d-i\infty}^{\d+i\infty} d\xi \, e^{-\f{1}{\xi}} (\xi s_{13} \bar s_{13} + \e)^{-2\bar h_1} (\xi s_{34} \bar s_{34} + \e)^{\bar h_1 +\bar h_2 - \bar h_3 - \bar h_4}  \\
\times (\xi s_{24} \bar s_{24}+\e)^{-\bar h_1 - \bar h_2 + \bar h_3 - \bar h_4}(\xi s_{23} \bar s_{23} + \e)^{\bar h_1 - \bar h_2 - \bar h_3 + \bar h_4}  H\(\tilde{u},\tilde{v}\)
\end{gathered} 
\ee

Following \cite{Melton:2023bjw} we now make a change of variable $ \xi=\d+i y $ to get,
\be
\begin{gathered}
\mathcal{C}_4(\s_i,\bar \s_i) = \f{\pi}{2} \int_{-\infty}^{\infty} dy \, e^{\f{i}{y-i\d}} (iy s_{13} \bar s_{13} + \e)^{-2\bar h_1}(iy s_{34} \bar s_{34} + \e)^{\bar h_1 +\bar h_2 - \bar h_3 - \bar h_4}  \\
\times (iy s_{24} \bar s_{24}+\e)^{-\bar h_1 - \bar h_2 + \bar h_3 - \bar h_4} (iy s_{23} \bar s_{23} + \e)^{\bar h_1 - \bar h_2 - \bar h_3 + \bar h_4} \\
\times  H\(\f{(iy s_{12} \bar s_{12})(iy s_{34} \bar s_{34} + \e)}{(iy s_{13} \bar s_{13} + \e)(iy s_{24} \bar s_{24} + \e)}, \f{\(iy s_{14} \bar s_{14}\)(iy s_{23} \bar s_{23}+\e)}{(iy s_{13} \bar s_{13} + \e)(iy s_{24} \bar s_{24} + \e)}\)
\end{gathered} 
\ee

where we have used $ (\xi s_{ij} \bar s_{ij}) = (i y s_{ij} \bar s_{ij} + \e) $ in the $ \d \to 0^+ $ limit. Now we break the integrals into two parts depending on $ y > 0  $ and $ y<0 $. This gives
\be
\begin{gathered}
\mathcal{C}_4(\s_i,\bar \s_i) = \f{\pi}{2} \int_{0}^{\infty} dy \, e^{\f{i}{y}} y^{-\bar h}e^{-i\f{\pi}{2}\bar h} ( s_{13} \bar s_{13} -i \e)^{-2\bar h_1}( s_{34} \bar s_{34} -i \e)^{\bar h_1 +\bar h_2 - \bar h_3 - \bar h_4} \\
\times ( s_{24} \bar s_{24}-i\e)^{-\bar h_1 - \bar h_2 + \bar h_3 - \bar h_4} ( s_{23} \bar s_{23} -i \e)^{\bar h_1 - \bar h_2 - \bar h_3 + \bar h_4} \\
\times H\(\f{( s_{12} \bar s_{12})( s_{34} \bar s_{34} -i \e)}{( s_{13} \bar s_{13} -i \e)( s_{24} \bar s_{24} -i \e)}, \f{\( s_{14} \bar s_{14}\)( s_{23} \bar s_{23}-i\e)}{( s_{13} \bar s_{13} -i \e)( s_{24} \bar s_{24} -i \e)}\)\\
 + \f{\pi}{2} \int_{0}^{\infty} dy \, e^{-\f{i}{y}} y^{-\bar h}e^{i\f{\pi}{2}\bar h} ( s_{13} \bar s_{13} + i \e)^{-2\bar h_1}( s_{34} \bar s_{34} + i\e)^{\bar h_1 +\bar h_2 - \bar h_3 - \bar h_4} ( s_{24} \bar s_{24}+i\e)^{-\bar h_1 - \bar h_2 + \bar h_3 - \bar h_4} \\
\times ( s_{23} \bar s_{23} + i\e)^{\bar h_1 - \bar h_2 - \bar h_3 + \bar h_4}   H\(\f{( s_{12} \bar s_{12})( s_{34} \bar s_{34} + i\e)}{( s_{13} \bar s_{13} + i\e)( s_{24} \bar s_{24} + i\e)}, \f{\( s_{14} \bar s_{14}\)( s_{23} \bar s_{23}+i\e)}{( s_{13} \bar s_{13} + i\e)( s_{24} \bar s_{24} + i\e)}\)
\end{gathered} 
\ee

where we have defined $ \bar h=\sum_{k=1}^4 \bar h_k $. By substituting $ y \to \f{1}{y} $ and performing the $ y $ integral we will get \eqref{main_res}.

To obtain the other scalar leaf amplitude we have to send $ \bar \s_i \to - \bar \s_i $ and use,
\be
(-x\pm i\e)^\D=e^{\pm i \pi \D}(x\mp i\e)^\D 
\ee
Then we have,
\be\label{sa_pc2}
\begin{gathered}
\mathcal{C}_4(\s_i,-\bar \s_i) = \f{i\pi}{2} \G\(\bar h - 1\) e^{-i\pi \bar h}( s_{13} \bar s_{13} - i \e)^{-2\bar h_1}( s_{34} \bar s_{34} - i\e)^{\bar h_1 +\bar h_2 - \bar h_3 - \bar h_4}  \\
\times ( s_{24} \bar s_{24}-i\e)^{-\bar h_1 - \bar h_2 + \bar h_3 - \bar h_4}( s_{23} \bar s_{23} - i\e)^{\bar h_1 - \bar h_2 - \bar h_3 + \bar h_4} \\ 
\times  H\(\f{( s_{12} \bar s_{12})( s_{34} \bar s_{34} - i\e)}{( s_{13} \bar s_{13} - i\e)( s_{24} \bar s_{24} - i\e)}, \f{\( s_{14} \bar s_{14}\)( s_{23} \bar s_{23}-i\e)}{( s_{13} \bar s_{13} - i\e)( s_{24} \bar s_{24} - i\e)}\) \\
- \f{i\pi}{2} \G\(\bar h - 1\) e^{i\pi \bar h} ( s_{13} \bar s_{13} +i \e)^{-2\bar h_1}( s_{34} \bar s_{34} +i \e)^{\bar h_1 +\bar h_2 - \bar h_3 - \bar h_4} ( s_{24} \bar s_{24}+i\e)^{-\bar h_1 - \bar h_2 + \bar h_3 - \bar h_4} \\
\times ( s_{23} \bar s_{23} +i \e)^{\bar h_1 - \bar h_2 - \bar h_3 + \bar h_4}   H\(\f{( s_{12} \bar s_{12})( s_{34} \bar s_{34} + i \e)}{( s_{13} \bar s_{13} + i \e)( s_{24} \bar s_{24} + i \e)}, \f{\( s_{14} \bar s_{14}\)( s_{23} \bar s_{23}+i\e)}{( s_{13} \bar s_{13} + i \e)( s_{24} \bar s_{24} + i \e)}\)
\end{gathered} 
\ee

\section{H-function}
\label{H_func}

$ H $-function and its properties can be found in \cite{Dolan:2000uw}. It is defined in terms of $ G $-functions by the following equation ( see (5.9) of \cite{Dolan:2000uw}), 
\be\label{rel2}
\begin{gathered}
     H(\a,\b,\g,\d;u,v)=\frac{\G(1-\g)}{\G(\d)} \G(\a)\G(\b)\G(\d-\a)\G(\d-\b) G(\a,\b,\g,\d;u,,1-v)
     \\  +\frac{\G (\g-1)}{\G(\d-2\g+2)} \G(\a-\g+1) \G(\b-\g+1)\G(\d-\g-\a+1)\G(\d-\g-\b+1) u^{1-\g} \\ \times G(\a-\g+1,\b-\g+1,2-\g,\d-2\g+2;u,1-v )
\end{gathered}
\ee

From Appendix \ref{appndxlc}, we know that,
\begin{equation}\label{abgd}
\begin{gathered} 
 \a= 2\bar h_1, \ \b= \bar h_1 + \bar h_2 - \bar h_3+\bar h_4, \ \g= 2 \bar h_1 + 2\bar h_2 - 1, \ \d = 2\bar h_1+ 2 \bar h_2
\end{gathered}
\end{equation}
i.e., $ \d=\g+1 $. For this special relation between $ \g $ and $ \d $, $ G $-functions can be written in terms of Hypergeometric functions using 
\be \label{rel3}
\begin{gathered}
     G(\alpha,\beta,\gamma,\gamma+1;u,1-v)=
\frac{1}{1-x-y}((1-y) \, _2F_1(\alpha-1,\beta-1;\gamma;x) \, _2F_1(\alpha,\beta;\gamma+1;1-y)\\
- x \, _2F_1(\alpha,\beta;\gamma+1;x) \, _2F_1(\alpha-1,\beta-1;\gamma;1-y))
\end{gathered}
\ee
where, $u=x(1-y)$ and $v=y(1-x)$.

\subsection{Scalar case}
\label{appB1}

In scalar case we have,
\begin{align*}
& \a= 1 + i\l_1 \\
& \b= 1+\f{1}{2}\(i\l_1+i\l_2-i\l_3+i\l_4\)\\
& \g= 1 + i\l_1 + i\l_2 \\
& \d = 2 + i\l_1 + i\l_2
\end{align*}

As mentioned in the main section we are interested in the configuration where $ \l_1=\l_2=\l_3=\l_4=\l $. Then $ H $-function is given by,
\be\label{rel4}
\begin{gathered}
     H(1+i\l,1+i\l,1+2i\l,2+2i\l;u,v) \\
     = -\f{\pi}{\sin\[2\pi i \l\]}\[\frac{\G(1+i\l)^4}{\G(1+2i\l)\G(2+2i\l)} G(1+i\l,1+i\l,1+2i\l, 2+2i\l;u,1-v) \right.   \\ 
      \left.   -\frac{\G(1-i\l)^4}{\G(1-2i\l)\G(2-2i\l)} u^{-2i\l} G(1-i\l,1-i\l,1-2i\l,2-2i\l;u,1-v) \]
\end{gathered}
\ee

On the support of $ \d(\l) $ we can expand this $ H $-function around $ \l=0 $ and keep upto $ \mathcal{O}(\l^0) $ terms only. Using \eqref{rel3} and the following expansion of Hypergeometric functions,
\be \label{rel11}
\begin{gathered}
 \, _2F_1(i\l,i\l;1+2i\l;x) = 1 + \mathcal{O}(\l^2)\\
 x \, _2F_1(1+i\l,1+i\l;2+2i\l;x) = -(1+2i\l) \log(1-x) - 2i\l \, \text{Li}_2(x) + \mathcal{O}(\l^2)
\end{gathered} 
\ee
one can show that,
\be \label{rel5}
\begin{gathered}
     G(1+i\l,1+i\l,1+2i\l,2+2i\l;u,1-v) =
\frac{1}{1-x-y}\[ -(1+2i\l) \log\(\f{y}{1-x}\) \right.\\
\left. -2i\l \{ \text{Li}_2(1-y) - \text{Li}_2(x)\}\] + \mathcal{O}(\l^2) \\
\end{gathered}
\ee

Using the above equation in \eqref{rel4} and, expanding around $ \l=0 $, we can get \eqref{Dbar1111}.


We know from \eqref{sim_uv}, that in our case, $ u, v $ are given by, $ u_{\pm} = z \bar z \pm (1+z \bar z)i\e $ and $ v_{\pm} = (1-z)(1-\bar z) \pm \{(1-z)(1-\bar z) - 1\}i\e $. Hence, we can write,
\be
x_{\pm}(1-y_{\pm}) =  z \bar z \pm (1+z \bar z)i\e, \ y_{\pm}(1-x_{\pm}) = (1-z)(1-\bar z) \pm \{(1-z)(1-\bar z) - 1\} i\e
\ee
The above equations can be solved for $ x_{\pm}, y_{\pm} $. The solutions are given by,
\be
x_{\pm} = z \mp \f{z^2+z-1}{\bar z - z}i\e, \  1-y_{\pm} = \bar z \pm \f{\bar z^2 + \bar z - 1}{\bar z - z}i\e
\ee

Let $ \zeta = \textnormal{sgn}\(\f{z^2+z-1}{\bar z - z}\), \, \bar \zeta = \textnormal{sgn} \left(\frac{\bar z^2 + \bar z - 1}{\bar z - z}\right) $.  Thus \eqref{Dbar1111} becomes,
\be
\begin{gathered}
\d(\l) H(1+i\l,1+i\l,1+2i\l,2+2i\l;(z \mp \zeta i\e)(\bar z \pm \bar \zeta i\e), (1-z \pm \zeta i\e)(1-\bar z \mp \bar \zeta i\e) ) \\
= \f{1}{\bar z - z \pm (\zeta + \bar \zeta )i\e}\[\{\log(z \mp \zeta i\e) + \log(\bar z \pm \bar \zeta i\e)\} \{\log(1-\bar z \mp \bar \zeta i\e) - \log(1-z \pm \zeta i\e) \} \right.\\
\left.  - 2 \text{Li}_2(z \mp \zeta i\e) + 2 \text{Li}_2(\bar z \pm \bar \zeta i\e) \]
\end{gathered} 
\ee

We take $ \zeta = \bar \zeta = +1. $ This can be achieved by letting $ z > \f{\sqrt{5}-1}{2}, \ \bar z > z $. However, in the region $ \f{\sqrt{5}-1}{2} < z < 1  $ we don't have any singularities for $ \log $ or $ \text{Li}_2 $ functions. Thus, in the limit $ z \to \bar z $, $ H $-function does not develop any simple pole. Hence, we take $ z > 1, \bar z > z $, i.e. $ z $ can approach $ \bar z $ from below. In this regime of $ z , \bar z $ we obtain \eqref{H-func-1}.




\subsection{Gluon case}
 
For MHV gluon scattering from \eqref{gl_wghts} and \eqref{abgd}, we have
\be
\a = 2+i\l_1, \ \b=2+\f{1}{2}(i\l_1+i\l_2-i\l_3+i\l_4) , \ \g=3+i\l_1+i\l_2, \ \d= 4+i\l_1+i\l_2
\ee

With $ \l_1=\l_2=\l_3=\l_4=\l $ one will get  the $ H $-functions appeared in \eqref{ga_pc1}.

\section{Tree level 4-point celestial amplitudes for massless scalars}
\label{scal_cel_amp}

The tree level momentum space 4-point amplitude is given by \eqref{s4pt}. In this section of the appendix, we compute the $ 4 $-point celestial amplitude in planar coordinates. As mentioned in the main section we take $ \e_1=\e_2=-1, \ \e_3=\e_4=+1 $. The momentum conserving delta function can be parametrized as,
\be\label{deltafn}
\begin{gathered}
\d^{(4)}\(\sum_{k=1}^4 p_k^\mu\) = \f{1}{4\om_4} \d\(\om_1 - \om_1^*\)\d\(\om_2 - \om_2^*\)\d\(\om_3-\om_3^*\)\d(r-\bar r)
\end{gathered} 
\ee
where
\be
\begin{split}
\om_1^* &= \om_4 \f{z_{24}\bar z_{34}}{z_{12} \bar z_{13}}, \ \om_2^* = \om_4 \f{z_{41}\bar z_{34}}{z_{12} \bar z_{23}}, \ \om_3^* = \om_4 \f{z_{24}\bar z_{41}}{z_{23} \bar z_{13}},\\
 \ r &= z_{12} z_{34} \bar z_{13} \bar z_{24}, \bar r = \bar z_{12} \bar z_{34} z_{13} z_{24} 
\end{split} 
\ee
The 4-point celestial amplitude is given by,
\be
\begin{gathered}
\mathcal{M}_4(\{z_i,\bar z_i, \bar h_i \}) = \prod_{k=1}^4 \int d\om_k \, \om_k^{2\bar h_k -1} A_4(p_1,p_2,p_3,p_4)\\
\end{gathered} 
\ee
where $ 2\bar h_k = \D_k = 1 + i\l_k $. Now, substituting $ A_4 $  from \eqref{s4pt} and using \eqref{deltafn} we get,
\be
\begin{gathered}
\mathcal{M}_4(\{z_i,\bar z_i, \bar h_i \}) 
= - \f{i \pi (2\pi)^4 \tilde \l}{2 z_{13} z_{24} \bar z_{13}\bar z_{24}} \Theta\(\f{z_{24}\bar z_{34}}{z_{12} \bar z_{13}}\)\Theta\(\f{z_{41}\bar z_{34}}{z_{12} \bar z_{23}}\)\Theta\(\f{z_{24}\bar z_{41}}{z_{23} \bar z_{13}}\) \d(z-\bar z) \\
\times \(\f{z_{24}\bar z_{34}}{z_{12} \bar z_{13}}\)^{2\bar h_1-1} \(\f{z_{41}\bar z_{34}}{z_{12} \bar z_{23}}\)^{2\bar h_2-1} \(\f{z_{24}\bar z_{41}}{z_{23} \bar z_{13}}\)^{2\bar h_3-1} \d(\b)
\end{gathered} 
\ee
where we have defined $ \b = \sum_{k=1}^4 \l_k $.  The cross ratios are given by,
\be z=\f{z_{12} z_{34}}{z_{13} z_{24}}, \ \bar z = \f{\bar z_{12} \bar z_{34}}{\bar z_{13} \bar z_{24}} \ee

Using the conformal symmetry we take three points to $ 0,1  $ and $ \infty $. More precisely we define an amplitude in the following way,
\be \begin{gathered}
\widetilde{\mathcal{M}_4}(z_,\bar z,\{\bar h_i \}) = \lim_{z_1, \bar z_1 \to \infty} (z_1 \bar z_1)^{2\bar h_1}\mathcal{M}_4( z_1,\bar z_1\to \infty, z_2=\bar z_2=1,  \\ \hspace{3cm} z_3=z,\bar z_3 = \bar z, z_4=\bar z_4=0, \{ \bar h_i \}) \end{gathered}
\ee

Remember that for scalars we have $ h_i = \bar h_i $. We obtain,
\be \label{scal_mell}
\begin{gathered}
\widetilde{\mathcal{M}_4}(z_,\bar z,\{\bar h_i \}) = \( - i (2\pi)^4 \tilde \l \) \f{\pi}{2} \Theta\(z-1\)  \d(z-\bar z) \ z^{2\bar h_1+ 2\bar h_2 - 2}  \(z-1\)^{2-2\bar h_2 - 2\bar h_3 } \d(\b)
\end{gathered}
\ee

If the imaginary part of the complex dimensions of all the scalars are same, i.e., $ \l_1=\l_2=\l_3=\l_4=\l $, then we have,
\be \label{scal_mells}
\begin{gathered}
\widetilde{\mathcal{M}_4}(z,\bar z, \l) = \( - i (2\pi)^4 \tilde \l \) \f{\pi}{8}  \Theta\(z-1\)  \d(z-\bar z) \, \d(\l)
\end{gathered}
\ee

\end{document}